\documentclass[manuscript,nonacm]{acmart}
\usepackage{fvextra}
\usepackage{multicol}
\usepackage{booktabs}
\usepackage{tabularx}
\usepackage{array}

\setcopyright{none}
\begin{document}

\title[Time Management Challenges and Strategies on r/ADHD]{r/ADHD: Community-Based Understandings of Time Management Challenges in Digital Environments}

\author{Shizhao Liu}
\email{sliu12@scu.edu}
\affiliation{%
  \department{Computer Science and Engineering}
  \institution{Santa Clara University}
  \city{Santa Clara}
  \state{California}
  \country{United States}
}

\author{Lei Cao}
\email{caolei@usc.edu}
\affiliation{%
  \department{Annenberg School for Communication and Journalism}
  \institution{University of Southern California}
  \city{Los Angeles}
  \state{California}
  \country{United States}
}

\author{Celia T Lee}
\email{lee04708@umn.edu}
\affiliation{%
  \department{Department of Family Social Science}
  \institution{University of Minnesota}
  \city{St.\ Paul}
  \state{Minnesota}
  \country{United States}
}

\author{Nikash Shanbhag}
\email{ngshanbhag@scu.edu}
\affiliation{%
  \department{Computer Science and Engineering}
  \institution{Santa Clara University}
  \city{Santa Clara}
  \state{California}
  \country{United States}
}

\author{Renkai Ma}
\email{mark@ucmail.uc.edu}
\affiliation{%
  \department{School of Information Technology}
  \institution{University of Cincinnati}
  \city{Cincinnati}
  \state{Ohio}
  \country{United States}
}

\author{Kai Lukoff}
\email{klukoff@scu.edu}
\affiliation{%
  \department{Computer Science and Engineering}
  \institution{Santa Clara University}
  \city{Santa Clara}
  \state{California}
  \country{United States}
}

\renewcommand{\shortauthors}{Liu et al.}

\begin{abstract}
  Community accounts of ADHD can complement clinical understandings by foregrounding members’ own definitions of time-management difficulties and support. We used human-led, LLM-assisted thematic analysis to examine 5,379 relevant r/ADHD posts from 2020 to 2025. Across five challenge and nine strategy themes, members described difficulties beginning intended tasks, disengaging from digital media, and sustaining support. Tools often demanded the self-regulation they were meant to support, while lost time and tool breakdown carried guilt and self-blame. Members drew on people, routines, and environments alongside digital tools, often seeking advice and recognition from peers. These accounts broaden the design problem to include the practical and emotional work of sustaining support. They motivate tools that reduce this burden and evaluations that consider recovery after distraction and emotional costs alongside task outcomes.
\end{abstract}



\maketitle

\section{Introduction}

Managing time in digital environments is difficult for many users. Smartphones and social media platforms are engineered to capture and hold attention, with designs that make time harder to track and regulate~\cite{roffarello2023defining, lukoff2021youtube}. For people with attention-deficit/hyperactivity disorder (ADHD), the challenge is even greater~\cite{ptacek2019clinical, panagiotidi2022attention, hariadi2026advancing}. Members of the online ADHD communities often use the term ``time blindness'' to describe difficulties in perceiving, estimating, and tracking time~\cite{tcherdakoff2026myself}. Prior work also identifies time management as a core challenge associated with ADHD~\cite{nakashima2022efficacy, ptacek2019clinical}. %
Indeed, prior studies have shown that ADHD symptoms are positively associated with problematic smartphone and internet use~\cite{panagiotidi2022attention, werling2022problematic}, and platforms such as TikTok and YouTube can intensify impulsive use and time blindness~\cite{khan2025inclusive, tcherdakoff2026myself}.
 
ADHD has been a fixture of public conversation for decades, from mass-media debates over its diagnosis and treatment~\cite{conrad2000hyperactive, conrad2014impending} to growing online information seeking about the condition across the 2000s and 2010s~\cite{zhao2022mental}. Public discussion of ADHD surged across social media including on platforms such as TikTok during the COVID-19 pandemic~\cite{leveille2024tell, tcherdakoff2026myself}. This growing public visibility and trends result in a broader rise in ADHD diagnoses and treatment-seeking after the pandemic~\cite{paul2025incidence}. One recent study found that during COVID-19, the prescriptions for both stimulant and non-stimulant ADHD medications increased significantly, with stimulants rising by 14\% and non-stimulants by 32\% compared with the pre-pandemic period~\cite{chai2024trends}. Although ADHD has long been studied in psychology and neuroscience, the ubiquity of digital technologies such as smartphones and social media has created environments that can actively worsen ADHD-related difficulties, particularly around managing time. A growing body of Human-Computer Interaction (HCI) research examines the digital challenges people with ADHD face~\cite{desrochers2019evaluation, khan2025inclusive, zhu2025focusview, hariadi2026advancing} and designs tools and strategies to address them~\cite{sonne2016framework, sonne2016mobero, ara2025alone, zhu2026scaffolding}, but the two are typically studied separately. Less is known about how community members themselves describe both their challenges and the strategies they try, and how the two connect. We believe this relationship could ground the design of self-regulation and time-management technologies in the community's own formulation of the problem.

A growing number of tools promise support for managing time in digital environments, from focus timers and reminder systems to distraction blockers. Reviews suggest that these tools can help users limit unwanted technology use and stay aligned with intended tasks~\cite{lyngs2019selfcontrol, roffarello2023achieving}. However, understanding everyday support requires looking beyond digital tools to encompass social and environmental resources. Furthermore, ADHD technology research has been critiqued for locating difficulties primarily within the person~\cite{spiel2022adhd}. Because clinical accounts and designer assumptions often diverge from what communities actually consider a problem~\cite{mankoff2010disability}, adults with ADHD frequently report struggling to find tools that fit how they work~\cite{desrochers2019evaluation}. Online communities provide opportunities to examine these difficulties in members’ own words. For example, accessibility and neurodiversity research on Reddit has examined collaborative practices among ADHD students~\cite{arnold2025beyond}, disclosure and workplace support among neurodivergent workers~\cite{newman2025disclosure}, and coping practices around reading difficulties and medication shortages~\cite{thach2025topic, shrestha2025adhd}. Yet whether the assumptions built into these tools match how the ADHD community describes its own digital time-management challenges remains largely unexplored. For example, which challenges members consider most pressing, and which strategies they pair with which challenges, would inform the design of ADHD-support technologies with insights that individual studies or tool evaluations often struggle to surface.
To address this gap, we analyze posts from r/ADHD, a community of approximately two million members~\cite{redditadhd}, to examine the time-management challenges within digital environments that community members describe, the strategies they propose for addressing them, and the relationships between them. Our study thus asks the following three research questions (RQs):

\begin{itemize}
    \item \textbf{RQ1:} What time-management challenges related to digital technology use are discussed within the r/ADHD community?
    \item \textbf{RQ2:} What strategies do r/ADHD community members discuss as potential ways to address these challenges?
    \item \textbf{RQ3:} What relationships exist between the time-management challenges and strategies discussed by r/ADHD community members?
\end{itemize}

\section{Background \& Related Work}

\subsection{ADHD and Time Management}\label{sec:adhd-time-management}

\citet{claessens2007review} define time management as behaviors aimed at achieving effective use of time while performing goal-directed activities. This definition was developed primarily in general work and educational contexts. In this paper, we extend this perspective to digital environments by examining how interactions with digital technologies can interfere with users' efforts to use time effectively in pursuit of their goals. 

In digital environments, maintaining such alignment is difficult for many users, even neurotypical individuals. Digital technologies interrupt the ongoing activity, displace the task a user intended to do, and make it hard to disengage once use has begun~\cite{stothart2015attentional, lukoff2018meaningful, tran2019modeling}, and many of these environments are deliberately engineered to capture and hold attention in exactly these ways. Prior work links push notifications to distraction~\cite{pielot2017productive}, smartphone use in bed to bedtime procrastination and poorer sleep~\cite{zhang2020effects}, and short-form, recommender-driven platforms such as TikTok to prolonged and repeated engagement~\cite{yang2025tiktok}. \citet{roffarello2023defining} identify eleven attention-capture deceptive designs, from infinite scroll to autoplay, that maximize the frequency and duration of use, and such features undermine users' sense of agency over their own time~\cite{lukoff2021youtube}.

For people with ADHD, these environments can present greater challenges, and research points to four relevant mechanisms. First, ADHD involves weakened inhibitory control, meaning difficulty suppressing prepotent responses and resisting interference~\cite{barkley1997behavioral}. Notifications and other digital cues may place additional demands on this control. Second, the dual pathway model identifies a motivational route involving a heightened preference for immediate reward and an aversion to delay~\cite{sonugabarke2003dual}. Digital media offers immediate rewards while intended tasks often pay off later. Third, ADHD involves difficulties estimating and tracking elapsed time~\cite{noreika2013timing}. Interfaces with infinite scroll and autoplay can further limit cues for stopping. Fourth, difficulties with emotion regulation are common in ADHD~\cite{shaw2014emotion}. These difficulties are relevant not only to digital media use but also to the upkeep that tools for managing time demand. Tools that require users to resist distractions, track time, and maintain routines may depend on the very regulation they aim to support. While these mechanisms have been studied largely in clinical and laboratory settings, our focus is on how people with ADHD describe these demands in everyday encounters with digital tools, the emotional costs involved, and how they arrange support around these difficulties.

\subsection{Designing for ADHD Time Management Challenges: From Research Prototypes to Everyday Tools}

Prior HCI research has proposed several approaches to support time management for people with ADHD. \citet{sonne2016framework} developed a design framework linking ADHD challenges to technological approaches, including activity structure, distraction reduction, and rewards. Examples include \textit{MOBERO}, which guides children through morning and bedtime routines using visual prompts, countdowns, and rewards~\cite{sonne2016mobero}, and a caregiver web application paired with a child’s smartwatch to structure study and break periods through prompts and gamified feedback~\cite{gonzalezgarcia2025smartpomodoro}. Other work draws on body doubling, which uses another person’s presence to support task initiation, focus, and completion~\cite{eagle2024body}. A virtual reality study reported faster task performance and better perceived focus with a body double~\cite{ara2025alone}. Recent participatory design work also explores how generative artificial intelligence (AI) might support task awareness, reflective task execution, and emotion regulation among students with ADHD~\cite{zhu2026scaffolding}.

Alongside support for tasks and routines, recent work examines how people with ADHD experience specific digital platforms. \citet{zhu2025focusview} designed \textit{FocusView}, which lets viewers customize visual and auditory elements in informational videos. Their evaluation found improved perceived viewability but differing preferences, with background music distracting some participants and supporting others. \citet{khan2025inclusive} examined difficulties with impulsive consumption and self regulation on platforms using personalized recommendations and proposed features supporting greater user control. \citet{mildner2025comparative} compared how people with and without ADHD recognize and avoid dark patterns on a mock social networking site and found that they were equally poor at recognizing dark patterns and manipulative interface designs but participants with ADHD are more often avoided engaging with them. Together, these studies offer design recommendations grounded in ADHD users' experiences of particular platforms and features.

Beyond research settings, people with ADHD also draw on commercial tools to support everyday functioning, including task management apps, reminder systems, focus timers, and distraction blockers~\cite{adda_tools}. For example, \textit{Focus Bear} is designed for people with ADHD and offers distraction blocking, daily routines, and break reminders~\cite{focusbear}. \textit{Skylight} provides a dedicated digital calendar for tracking tasks and appointments~\cite{skylight}. Yet evidence for the sustained effectiveness of digital self control tools remains limited~\cite{roffarello2023achieving}. Adults with ADHD report struggling to find tools that fit how they work~\cite{desrochers2019evaluation}, and ADHD technology research has been critiqued for prioritizing behavioral normalization over users’ own goals~\cite{spiel2022adhd}. Broader digital well being research also identifies limitations of restrictive tools. Commitment interfaces let users limit future access to distracting apps or websites, reducing reliance on immediate self control, but these restrictions can also block content users need~\cite{lukoff2023switchtube}. An alternative is to adjust features within an app to support users’ intentions~\cite{lukoff2021youtube}. \textit{SwitchTube} explored this approach by allowing users to toggle between modes that prioritize search or recommendations. Participants rated the switchable version higher in sense of agency and goal alignment than the version prioritizing recommendations~\cite{lukoff2023switchtube}. However, one potential gap is a mismatch between mechanism and difficulty since these tools primarily rely on external support such as reminders, timers, or task breakdowns to scaffold users' behavior, designs that inherit a problem formulation the community they serve may not share.

Community knowledge offers one path toward closing this gap. Body doubling illustrates how practices used within neurodivergent communities can inform research~\cite{eagle2024body} and the design of new tools, including virtual reality prototypes~\cite{ara2025alone}. This motivates systematically examining how people with ADHD describe their time-management challenges and share strategies for managing them. Such accounts can identify needs and approaches that existing tools overlook, grounding design in members’ everyday experiences.

\subsection{Towards a Community-Based Understanding of Time Management Challenges}

ADHD technology research has been critiqued for framing difficulties as individual deficits to be corrected~\cite{spiel2022adhd}. Yet community accounts may not share this formulation. In assistive technology research, researchers and community members may differ in which problems deserve attention and what forms of support are appropriate~\cite{mankoff2010disability}. Applied to ADHD tools, this gap could mean designing to reduce screen time when members are more concerned with the emotional aftermath of lost time or the upkeep a tool requires. Relatedly, digital wellbeing scholarship questions placing responsibility on individuals alone and draws attention to interfaces and broader social conditions~\cite{docherty2022repoliticizing, roffarello2025heuristics}. Reading community accounts alongside clinical research helps us examine the demands of using digital tools rather than locating the problem solely in the user.

Whether tool mechanisms match the challenges people with ADHD describe remains an empirical question. A community perspective can reveal which digital time management challenges members discuss, what strategies they propose, and how challenges and strategies are related in their accounts. In turn, these accounts can reveal where existing design frameworks address members’ concerns and where they may need to be extended~\cite{sonne2016framework, lyngs2019selfcontrol}.

We ground our analysis in r/ADHD, the largest subreddit dedicated to individuals with ADHD~\cite{redditadhd}. Online ADHD communities offer a window onto how ADHD is experienced and managed in everyday life, surfacing struggles and coping practices that rarely appear in clinical or lab-based studies~\cite{eagle2023cant, ginapp2023experiences}. Reddit is particularly well suited to examining such experiences in depth. Its pseudonymous, long-form, community-based structure supports candid self-disclosure around stigmatized mental-health topics~\cite{dechoudhury2014mental}, and prior work in both computational and qualitative fields has demonstrated that Reddit discussions surface the challenges people with ADHD encounter and the strategies they develop in response~\cite{arnold2025beyond, thach2025topic, zhang2025seeking}.

\section{Methods}\label{sec:methods}
 
To analyze the corpus of r/ADHD posts, we used a human-led workflow assisted by a large language model (LLM), which enabled us to draw upon established methods for thematic analysis while scaling them up to analyze a large volume of posts. In particular, we used a thematic analysis approach called template analysis, a codebook-based approach to qualitative analysis in which researchers develop an initial coding template from a subset of the data, iteratively refine it through successive applications, and then apply the finalized template to the full dataset~\cite{brooks2015utility, king2012doing}. Unlike approaches in which themes emerge throughout the analysis, template analysis fixes a structured, hierarchical codebook partway through, making it well suited to team-based coding, reliability assessment, and large-scale annotation. In this workflow, human researchers retained responsibility for interpretive decisions, including code creation, code boundary refinement, theme development, and adjudication of ambiguous cases.

Our analysis draws on two strands of prior work. Within HCI, researchers have paired computational and qualitative methods to study social media, including statistical characterizations of Reddit mental health communities~\cite{dechoudhury2014mental}, topic models guiding close reading of parenting subreddits~\cite{ammari2018pseudonymous}, and qualitative and statistical analyses of Instagram depression disclosures and responses~\cite{andalibi2017sensitive}. A second strand examines computational support for qualitative analysis, including comparisons of topic modeling and grounded theory~\cite{baumer2017comparing}, computational thematic analysis tools~\cite{gauthier2022computational}, and researchers’ preferences for AI support~\cite{feuston2021putting, jiang2021supporting}. Outside HCI, \citet{nelson2020computational} formalizes this division of labor as computational grounded theory, combining computational pattern detection, qualitative refinement, and computational confirmation. Our workflow similarly combines computation and human interpretation but begins with template analysis. Human researchers developed and froze the codebook, the LLM applied it at scale with reliability assessed as described in (Sections~\ref{sec:prompt} and~\ref{sec:reliability}), and researchers returned to close reading for interpretation (Section~\ref{sec:closereading}). Annotation counts describe code prevalence within the filtered corpus and are reported in tables, while the findings narrative is organized by analytic salience to the research questions.

\begin{figure*}[t]
  \centering
  \includegraphics[width=0.9\textwidth]{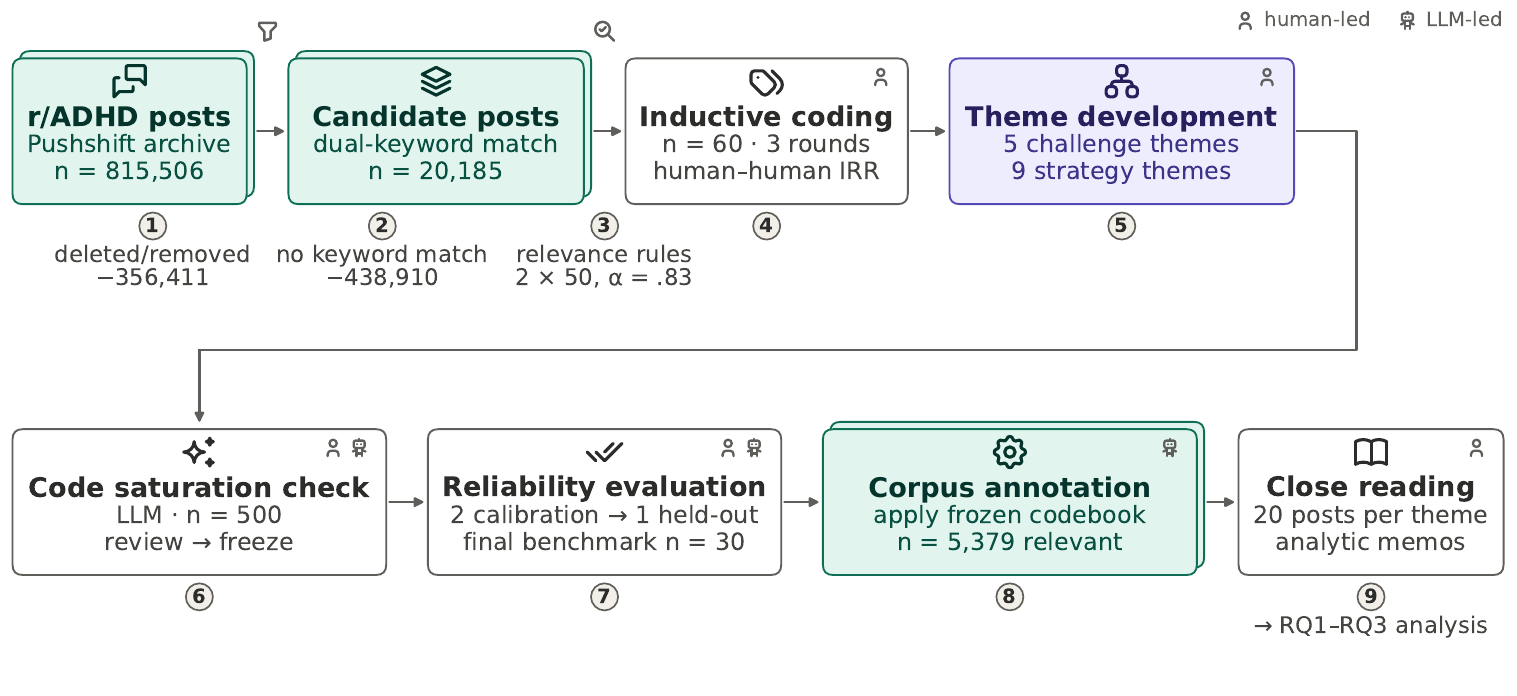}
  \caption{Overview of the analysis pipeline. Phases 1--3 cover data collection, keyword filtering, and relevance screening (Sections~\ref{sec:dataset} and~\ref{sec:screening}); phases 4--5 are the human-led codebook and theme development (Sections~\ref{sec:codebook} and~\ref{sec:themes}); phases 6--8 are the LLM-assisted saturation check, reliability evaluation, and annotation of relevant posts (Sections~\ref{sec:prompt} and~\ref{sec:reliability}); phase 9 is the close reading that feeds the RQ1--RQ3 analysis (Section~\ref{sec:closereading}). Icons mark whether each phase was human-led, LLM-led, or both.}
  \Description{A nine-phase left-to-right flowchart. phase 1: r/ADHD posts from the Pushshift archive, n = 815,506; deleted or removed posts (356,411) are dropped, leaving 459,095. phase 2: dual-keyword matching yields n = 20,185 candidate posts for relevance screening. phase 3: relevance screening with two calibration rounds of 50 posts. phase 4: inductive coding of 60 posts over three rounds with human--human IRR. phase 5: theme development producing 5 challenge themes and 9 strategy themes. phase 6: an LLM-based code saturation check on 500 posts, followed by review and codebook freeze. phase 7: reliability evaluation across three rounds, with a final held-out benchmark of 30 posts. phase 8: annotation of n = 5,379 relevant posts using the frozen codebook. phase 9: close reading of 20 posts per theme with analytic memos, feeding the analysis for RQ1 through RQ3. Icons mark each phase as human-led, LLM-led, or both.}
  \label{fig:pipeline}
\end{figure*}

The full workflow proceeded in six phases: relevance screening (Section~\ref{sec:screening}), human codebook development (Section~\ref{sec:codebook}), theme development (Section~\ref{sec:themes}), LLM prompt development and codebook saturation check (Section~\ref{sec:prompt}), reliability evaluation and LLM-assisted annotation of relevant posts (Section~\ref{sec:reliability}), and close reading and interpretation (Section~\ref{sec:closereading}). Figure~\ref{fig:pipeline} summarizes the full pipeline as nine stages, from data collection through close reading.

\subsection{Dataset}\label{sec:dataset}
 
We collected data from the subreddit r/ADHD using the Reddit archive available via Academic Torrents~\cite{academictorrents}, which compiles posts (top-level submissions to the subreddit) and comments (replies to those posts). 
This timeframe, between January 1, 2020, and November 8, 2025, was selected to encompass two major contextual shifts that likely reshaped ADHD-related discourse online, the COVID-19 pandemic and the ADHD trend on TikTok. The initial dataset contained 815,506 posts. We excluded posts marked as ``[deleted]'' or ``[removed]'', leaving 459,095 posts. Because our focus is on posts that address time management in a digital context, we created one keyword list for time management (with keywords such as ``procrastination'', ``planning'', and ``deadlines'') and one for digital environments (e.g., ``social media'', ``gaming'', and ``smartphone''). Appendix Table~\ref{tab:keywords} provides the full keyword lists. Only posts with at least one term from each of these keyword lists were retained, yielding 20,185 candidate posts for relevance screening.
 
\subsection{LLM-Assisted Thematic Analysis Pipeline}\label{sec:pipeline}
 
\subsubsection{Relevance screening}\label{sec:screening}

After filtering, three researchers removed deleted entries, spam, promotional content, posts not written in English, and posts containing only a title, link, or image. Because keyword filtering still retrieved irrelevant posts, we conducted a relevance coding pass before codebook development (see Appendix~\ref{app:relevance} for the rules and procedure). %
Posts were included only if they described a time management challenge caused by or directly shaped by digital technology use. For example, we excluded a post focused on ADHD medication effectiveness, dosage changes despite references to ``emails,'' ``screen time,'' and Reddit ``threads,''. We also excluded posts describing strategies without a challenge because we could not establish the difficulty being addressed or its relation to digital technology.

Three researchers independently coded relevance (relevant = 1, not relevant = 0) across two blind calibration rounds of 50 posts each, refining the rules after each round until agreement exceeded .80 (Krippendorff’s $\alpha = [0.83]$)~\cite{krippendorff2019content}. Remaining disagreements were resolved through consensus discussion.

\subsubsection{Human codebook development}\label{sec:codebook}

The same three researchers developed a codebook for the LLM analysis using a random sample of 60 posts, following template analysis of combining \emph{a priori} codes with codes developed from the data~\cite{brooks2015utility, king2012doing}. Challenge codes were developed inductively through open coding. Strategy codes began with categories from prior work on digital self control tools~\cite{lyngs2019selfcontrol}, which we retained, revised, or discarded based on the data, alongside additional inductive codes. The posts were randomly divided into three batches of 20, with one batch coded in each round. All three researchers independently coded each batch before discussing discrepancies and refining the codebook. The first round established the initial codebook, while the next two calibrated coders and refined definitions. After each round, we calculated inter-rater reliability (IRR) among human coders to identify unclear codes and track agreement.

In developing codes, we prioritized experiences connected to ADHD rather than general descriptions of digital distraction. Challenge codes included \emph{autopilot app-opening}, or opening apps without a conscious decision, \emph{can't re-enter focus afterward}, or difficulty regaining focus after interruption, and \emph{guilt about resting}, or feeling unable to relax while tasks remain unfinished. Strategy codes included \emph{separate accounts for work and entertainment} and \emph{accountability partner}, capturing changes to the working environment and support from another person.

\subsubsection{Theme development}\label{sec:themes}

Three researchers developed themes from the finalized human codebook before LLM annotation, following codebook thematic analysis~\cite{braun2022thematic}, specifically template analysis~\cite{brooks2015utility, king2012doing}. The team grouped codes by conceptual relatedness and close reading of members’ posts, then refined boundaries through discussion to ensure coherent and distinct themes. For example, \emph{guilt over lost time} and \emph{guilt about resting} formed the subtheme \emph{guilt over time perceived as misspent} within the theme \emph{emotional consequences of time loss}.

We distinguished \emph{digital disengagement failure}, or difficulty ending a session without a specific task in mind, from \emph{digital task avoidance}, where digital use displaced an intended activity. This distinction reflects the emphasis on intended goals in our definition of time management~\cite{claessens2007review}. We also placed \emph{failed productivity-tool attempts} under the challenge theme \emph{self-regulation strategy breakdown}, since this code captured barriers encountered when strategies were tried and abandoned, rather than the strategies themselves.

\subsubsection{LLM prompt development and codebook saturation check}\label{sec:prompt}

To prepare for annotation at scale, we iteratively refined structured prompts (Appendix~\ref{app:prompts}). %
Before freezing the codebook, we assessed code saturation, the point at which no new codes are identified~\cite{hennink2017code}, using an additional 500 posts. We used the LLM pipeline to screen a larger sample, building on prior work on deductive coding at scale~\cite{ma2026negotiating, xiao2023supporting, meng2025deconstructing, chew2023llm}. The prompts instructed the LLM to flag content not covered by the existing codebook for human review.

The code saturation assessment of the additional 500 posts yielded 10 candidate codes (4 challenge and 6 strategy). Three researchers reviewed all candidates in team meetings, comparing them with existing codes and their mapping to themes. Most overlapped with existing codes and were recategorized through discussion. For example, the proposed code \textit{journal reflection} was considered similar to \textit{keep track of time spent on tasks}. Remaining candidates were excluded because they did not clearly connect to digital time management challenges. These included medication treatment (6 posts), other forms of medical support, and advance meal preparation, consistent with (Section~\ref{sec:screening}). The review yielded no additional codes or changes to subthemes or themes, providing evidence of code saturation within the study’s scope. We therefore finalized the template, following template analysis~\cite{brooks2015utility, king2012doing}, and proceeded to annotation of relevant posts (Section~\ref{sec:reliability}).

\subsubsection{Reliability evaluation and LLM-assisted annotation of relevant posts}\label{sec:reliability}
 
To validate the codebook and LLM annotation before scaling up, we assessed human--LLM IRR in 3 rounds after the codebook saturation check. In each round, we drew a fresh held-out sample of posts ($N = 30$) that had not been used in codebook development or prompt refinement. Three coders independently coded the sample, treating each post as a single predefined coding unit and applying each challenge and strategy code with multiple codes allowed per post. We then ran the current prompt over the same posts and computed per-code human--LLM agreement against this reference using Krippendorff's $\alpha$. Codes with low agreement were addressed between rounds by clarifying code definitions and decision rules in the prompt or, where a code could not be applied reliably, by merging or excluding it, and the revised prompt was re-tested on the next round's fresh sample. On the final 30-post benchmark, agreement over post-level theme sets, with the LLM included as a fourth coder, was $\alpha = .81$ for challenges and $.72$ for strategies (Table~\ref{tab:irr}). Four of fourteen themes met our reporting rule of at least three posts where a majority of human coders applied the theme (Table~\ref{tab:irr-themes}). For those four themes, human--LLM agreement ranged from $.64$ to $.90$ and human--human agreement from $.76$ to $.93$.

Following reliability assessment, we applied the finalized screening and annotation pipeline (Appendix~\ref{app:prompts}) to the keyword-matched candidate posts using GPT-5.4, consistent with recent HCI work on LLM coding at scale~\cite{gao2024coaicoder, ma2026negotiating, wang2024humanllm, xiao2023supporting}. The pipeline retained 5,379 relevant posts and generated JSON annotations for these posts using the frozen codebook (Appendix~\ref{app:codebook}), with subthemes and themes assigned through the code-to-theme mapping.

\begin{table}
  \begin{tabular}{lcc}
    \toprule
    & Human--human & Human--LLM \\
    \midrule
    Challenge themes & .88 & .81 \\
    Strategy themes  & .80 & .72 \\
    \bottomrule
  \end{tabular}
  \medskip
  \caption{IRR on the held-out benchmark sample ($N = 30$ posts;
  Krippendorff's $\alpha$ with MASI distance over post-level theme
  sets). Human--human $\alpha$ is computed over three independent
  coders; human--LLM $\alpha$ treats the pipeline as a fourth coder
  and was computed prior to adjudication.}
\label{tab:irr}
\end{table}

\begin{table}
  \centering
  \begin{tabular}{lccc}
    \toprule
    Theme & Support & Human--human & Human--LLM \\
    \midrule
    \multicolumn{4}{l}{\emph{Challenge themes}} \\
    \quad Digital task avoidance & 6 & .93 & .90 \\
    \quad Emotional consequences & 5 & .92 & .88 \\
    \quad Digital disengagement failure & 3 & .86 & .64 \\
    \quad Task-embedded digital interruption & 1 & --- & --- \\
    \quad Self-regulation strategy breakdown & 0 & --- & --- \\
    \addlinespace
    \multicolumn{4}{l}{\emph{Strategy themes}} \\
    \quad Social co-regulation & 3 & .76 & .67 \\
    \quad Digital friction engineering & 1 & --- & --- \\
    \quad Other strategy themes (7) & 0 & --- & --- \\
    \midrule
    \emph{All challenge themes (set-based)} & & .88 & .81 \\
    \emph{All strategy themes (set-based)} & & .80 & .72 \\
    \bottomrule
  \end{tabular}
  \medskip
  \caption{Reliability by theme on the held-out benchmark
  ($N = 30$ posts). Per-theme values are binary presence/absence
  Krippendorff's $\alpha$; support is the number of posts where a
  majority of human coders applied the theme, and $\alpha$ is
  reported for support $\geq 3$. Aggregate rows use set-valued
  $\alpha$ with MASI distance over full theme sets. Human--LLM
  treats the pipeline as a fourth coder, computed prior to
  adjudication.}
  \label{tab:irr-themes}
\end{table}

\subsubsection{Close reading and interpretation}\label{sec:closereading}

The pipeline retained 5,379 relevant posts, which formed the analytical corpus. We used annotations from these relevant posts as a retrieval index for qualitative interpretation rather than as findings in themselves. For each theme, we sampled 20 tagged posts, stratified by subtheme to include less prevalent subthemes, and added posts near code boundaries. Three researchers read the samples and wrote an analytic memo for each theme~\cite{saldana2021coding}, documenting its central claim, subtheme relationships, boundaries with adjacent themes, negative cases, and candidate quotations. In team meetings, we used these memos to organize the findings narrative. Reliability evaluation and prevalence reporting remained at the level of the frozen codebook. For reporting, we paraphrased all quotations, checked that searches for the paraphrases did not retrieve source posts, and omitted post identifiers.

\subsection{Ethical Considerations}\label{sec:ethics}

Our data consist of publicly posted submissions to r/ADHD obtained from the Pushshift-derived Reddit archive distributed via Academic Torrents~\cite{baumgartner2020pushshift, academictorrents}, which we accessed for non-commercial research consistent with Reddit's Public Content Policy. The study analyzes archival public data and involves no intervention or interaction with individuals, and we collected no usernames, profile information, or user posting histories; it therefore does not meet the federal definition of human subjects research in the United States (45 CFR 46.102(e)). Public availability does not eliminate privacy interests, particularly in a community where members discuss stigmatized health experiences pseudonymously~\cite{proferes2021studying, franzke2020internet}. We took four steps to respect those interests. We excluded all posts that authors had deleted or moderators had removed prior to archiving. We report prevalence only in aggregate, at the level of themes, and present individual accounts solely as paraphrased excerpts. We paraphrase all excerpts rather than quoting verbatim to reduce the risk of tracing excerpts to their source posts~\cite{ayers2018dont, bruckman2002studying, markham2012fabrication}, and we verified that each paraphrased excerpt does not retrieve its source post via search engines and that the source post remained publicly visible at the time of writing. Finally, posts were annotated with GPT-5.4 via the OpenAI API, whose terms exclude use of submitted data for model training by default;\footnote{OpenAI, ``Enterprise privacy at OpenAI,'' https://openai.com/enterprise-privacy/, accessed September 2026.} the corpus is stored on access-restricted university systems and is not redistributed.

\section{Results}\label{sec:results}
\subsection{RQ1 - What time-management challenges related to digital technology use are discussed within the r/ADHD community?}\label{sec:RQ1}

\begin{figure}[htbp]
    \centering
    \includegraphics[width=0.7\textwidth]{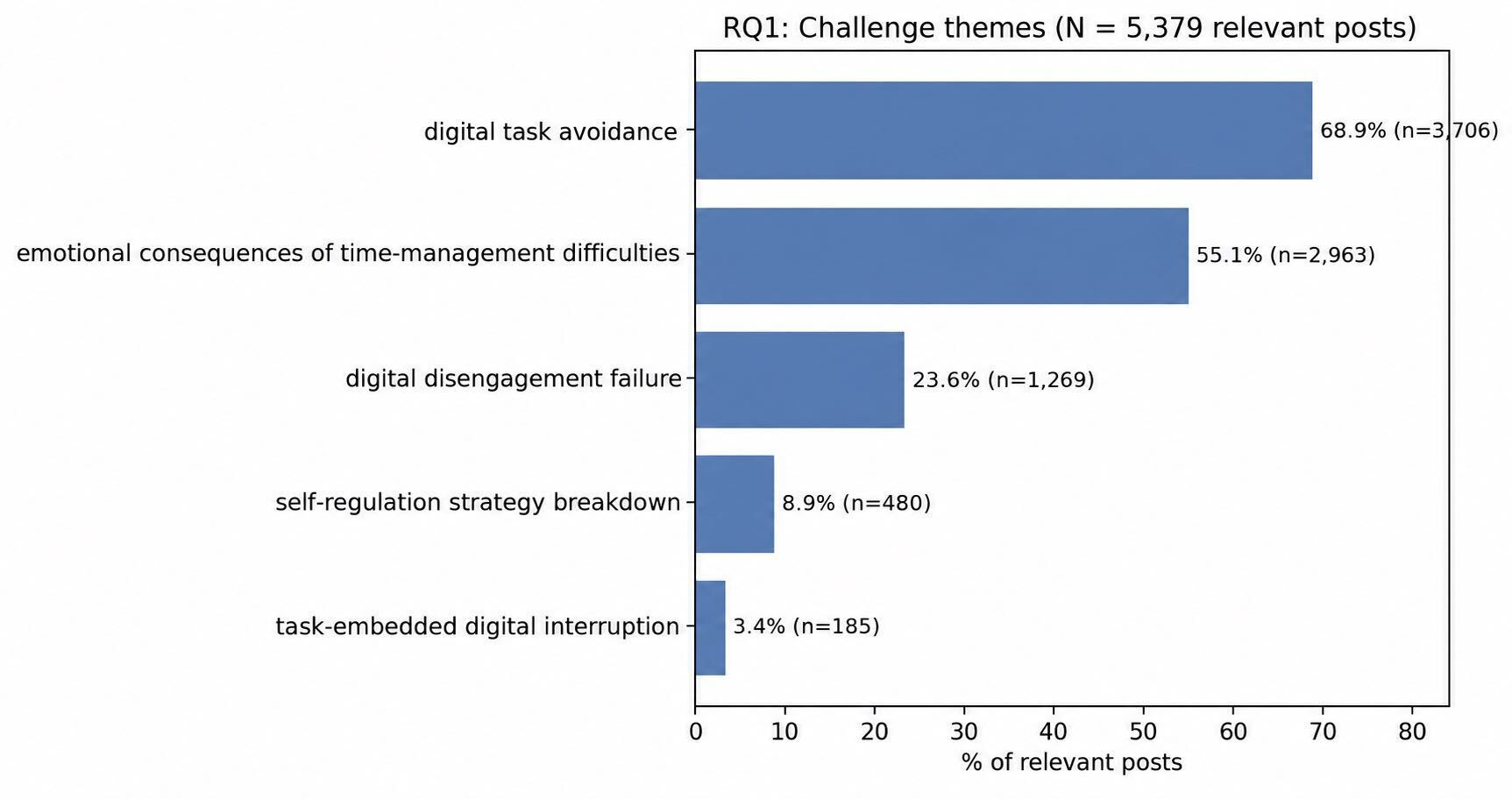}
    \caption{Distribution of challenge themes across N = 5,379 relevant posts. Bars show the percentage of relevant posts coded for each theme (n in parentheses); themes are not mutually exclusive, so percentages sum to more than 100\%.}
    \label{fig:rq1}
\end{figure}

The codebook comprises five challenge themes, and Figure~\ref{fig:rq1} shows the prevalence of all five across the relevant posts. We report four overarching themes characterizing the digital time-management challenges discussed by members of r/ADHD below. The fifth, \textit{task-embedded digital interruption}, was the least prevalent (185 posts, 3.4\%), so we note it here rather than giving it a separate subsection. Each subsection is labeled with the codebook theme plus a one-line summary of the close-reading finding. 
Members describe digital activity put off tasks they know they intended to do until urgency forces a start (Section~\ref{sec:RQ1-1}), remain in sessions with no intended tasks whose length they discover only afterward (Section~\ref{sec:RQ1-2}), watch the tools they adopt in response fail through fading novelty and add to the feeling of guilt (Section~\ref{sec:RQ1-3}), and carry the accumulated emotional consequence as evidence against themselves rather than the tools (Section~\ref{sec:RQ1-4}).

\subsubsection{Digital task avoidance: putting off tasks until they feel urgent}\label{sec:RQ1-1}

In the corpus’s most prevalent challenge theme, the problem is not that members forget the task, but rather they describe scrolling with the deadline in view, waiting for a feeling of urgency that compels them to start. Digital task avoidance captures members turning to digital activities in place of a task they intended to do ($N = 3,706$, or 68.9\% of relevant posts).
Members attribute their inability to complete assignments, work, chores, and get adequate sleep to their use of phones and video games. One member shared that \textit{“I’ll be well aware that I have a deadline for an assignment coming up, but instead I’ll just scroll through my phone while still feeling overwhelmed by the thought of doing my assignments”}. This avoidance extends to the bedtime routine as well, with one member saying \textit{“I would come home from work and lay in bed scrolling through my phone for hours, knowing I need to sleep for the next work day but I couldn’t”}.
Members’ accounts suggest that the primary barrier was not the device itself, but in the cognitive and emotional states that shaped its use, including difficulty initiating tasks without urgency, racing thoughts, and feelings of overwhelm that scrolling temporarily relieved. First, members describe needing a sense of urgency before they can start a task, and until a deadline is close enough to supply it, they put the task off: \textit{“I need a sense of urgency to get things done. If there is no urgency, I don’t get it done”}. In the meantime, members defer the task through an inner negotiation: \textit{“I’ll love to work on this new skill, but maybe later”}. Second, some supplied their experiences, saying \textit{“to me my head feels like the Internet, always on, always running”}. Another member describes that ADHD feels like \textit{“a wall that stops me from doing what I need to do”}. Lastly, members shared using screen time to regulate their emotions related to task avoidance. One member turned to feeds to \textit{“avoid stress”} at work. Another member shared that \textit{“I would scroll while still feeling overwhelmed by the thought of doing my assignments”}.

\subsubsection{Digital disengagement failure: losing track of time until hours pass}\label{sec:RQ1-2}

\textit{Digital disengagement failure} begins with no particular task in mind. Members often describe screen use as a default state rather than a chosen activity, and report that the time lost only became apparent after it was gone. It captures members’ accounts of screen sessions they could not end such as scrolling feeds, watching videos, and playing games that continued without a clear purpose or stopping point. It also captures the phone opened automatically in idle moments before any task has formed ($N = 1,269$, or 23.6\% of relevant posts).
One member intended to scroll for \textit{“5 mins”}, only to realize that \textit{“3 hours passed”}. Another described that \textit{“I’ll have this period of 5 minutes that I don’t know what to do with and I end up on my phone doomscrolling or hitting a dopamine farm like TikTok”}. These accounts suggest that unstructured time could become extended use before members noticed how much time had passed. Members consequently described digital media use as a state from which it was difficult to transition. One characterized browsing as \textit{“the default, ground state,”} while beginning anything important required \textit{“an unbearable amount of willpower and energy”}. Entry, by contrast, could occur automatically, as another explained, \textit{“My muscle memory took over and within seconds I am trapped”}. When they want to exit digital media, the state did not hold: \textit{“I will almost instinctively open up Twitter/Facebook/Youtube/Reddit repeatedly, closing them down when I realize what I’m doing and only to reflexively open up a new tab again”}. Sessions could also continue beyond enjoyment, with one member remaining on screens \textit{“until I’m not even having fun”}. These accounts linked difficulty noticing elapsed time with difficulty of stopping or leaving a protective state.

\subsubsection{Self-regulation strategy breakdown: tools stop working and add to the guilt}\label{sec:RQ1-3}

The strategies members built to improve self-regulation often failed. \textit{Self-regulation strategy breakdown} captures accounts of tools and routines that lost effectiveness, fell out of use, or were abandoned ($N = 480$, 8.9\% of relevant posts). Members described fading novelty, discouraging feedback, and upkeep that competed with intended work. Tools could also require the regulation they were intended to support.

Usefulness often diminished as tools became familiar. Members described methods that \textit{“would help me a bit on the first try but would fail to help literally on the second try”} and a planner forgotten \textit{“after about 2 weeks”}. When reminders lost their novelty, one member explained, \textit{“I started ignoring them and now I don’t even want to look at them”}. Tools also depended on members enforcing their own rules. One member described abandoning goals \textit{“as soon as I’m mentally aware that there are no consequences for missing it”}. Tool failure therefore emerged not as a single breakdown, but as a process through which novelty faded, feedback acquired emotional costs, maintenance competed with action, and external support reverted to self-enforcement ultimately.

On top of it, feedback could itself become something to avoid. One member described anxiety caused by a to-do app’s red notification badge, explaining that \textit{“I end up muting them or ignoring/avoiding them altogether and then forgetting everything”}. Repeated dismissal could also undermine confidence in a strategy beyond a particular tool. After years of silencing phone alarms, another member worried they had \textit{“just completely burned out that strategy”}.

\subsubsection{Emotional consequences of time-management difficulties: lasting guilt and self-blame over lost time}\label{sec:RQ1-4}

The second most prevalent theme centered around the emotional consequences of missed tasks and abandoned tools ($N = 2,963$, 55.1\% of relevant posts). Members described exhaustion, helplessness, persistent guilt, and self-directed anger associated with repeated self-regulation difficulties.

Members described emotional exhaustion from repeated self-regulation struggles and unsuccessful attempts to sustain strategies (Section~\ref{sec:RQ1-3}). One member shared that \textit{“I feel like everyone with ADHD understands the feeling that you're ALWAYS struggling with something”}. When under accumulating demands, \textit{“my head feels like a jumble of a million wires that I’m frantically trying to untangle”}. When there’s no way out, exhaustion becomes helplessness about bad digital habits, and they express that \textit{“I’m so tired of living through this cycle but I can’t seem to break out of it on my own”}. In many of these accounts, the self-efficacy is specifically about direction. Members no longer know what to try. It appears at the level of the task: \textit{“I really want to get all of these tasks accomplished but I just don't know how to get started”}. When members’ own ideas run out, they will turn to the community, a strategy we return to in Section~\ref{sec:RQ2-1}: \textit{“Does anyone know what I can try to just do anything apart from just sitting there doing nothing or scrolling through social media? I just don’t know what to do anymore”}. 

Members described guilt and shame as both consequences of time management difficulties and feelings that further disrupted focus. One member explained that \textit{“the guilt and shame is crushing at times. It only makes it harder to focus”}. Another stated that \textit{“Worrying too much about wasting my time ironically results in me wasting my time the most”}. \textit{Guilt over time perceived as misspent} also persisted despite explanations for these difficulties. As one member shared, \textit{“Even now that I know the rational answer for why those hours slip by, I still get pangs of guilt”}. These concerns extended beyond lost time to judgments about personal potential and character. One member wrote, \textit{“I am a smart person and a good student. That is a fact. I just never get to reach my full potential”}. Others expressed anger toward ADHD, stating that \textit{“honestly I hate having adhd. I feel like it’s ruining my life”}, or questioned their character, suggesting that \textit{“maybe I’m just lazy and irresponsible as a person”}. These accounts suggest how frustration with time management difficulties became negative judgments about the self, while the resulting emotions made those difficulties harder to address.

Together, these accounts indicate that digital time-management struggles carry an emotional aftermath that is itself part of the challenge. The helplessness members describe emerges when their attempts to control their digital time repeatedly fail, leaving them uncertain about how to change their behavior. They adopt apps to manage their digital time, see those attempts fail in turn, and are left uncertain about what to do next. As one member put it: \textit{“even if I did manage ninety straight days of self-discipline, I’d still end up right where I started... Maybe the problem isn’t the system or the struggle, it's me”}. Another complaint is a focus timer that \textit{“mostly just highlights how much time I wasted”}. Further, members located the weakness in the tool’s dependence and their lost agency on themselves. For example, a brain that \textit{“takes issue with having goals already and plans are an insult to injury”}. For this community, managing digital time also meant preserving agency so that each loss of time is evidence against the self.

\subsection{RQ2 - What strategies do r/ADHD community members discuss as potential ways to address these challenges?}\label{sec:RQ2}

\begin{figure}[htbp]
    \centering
    \includegraphics[width=0.7\textwidth]{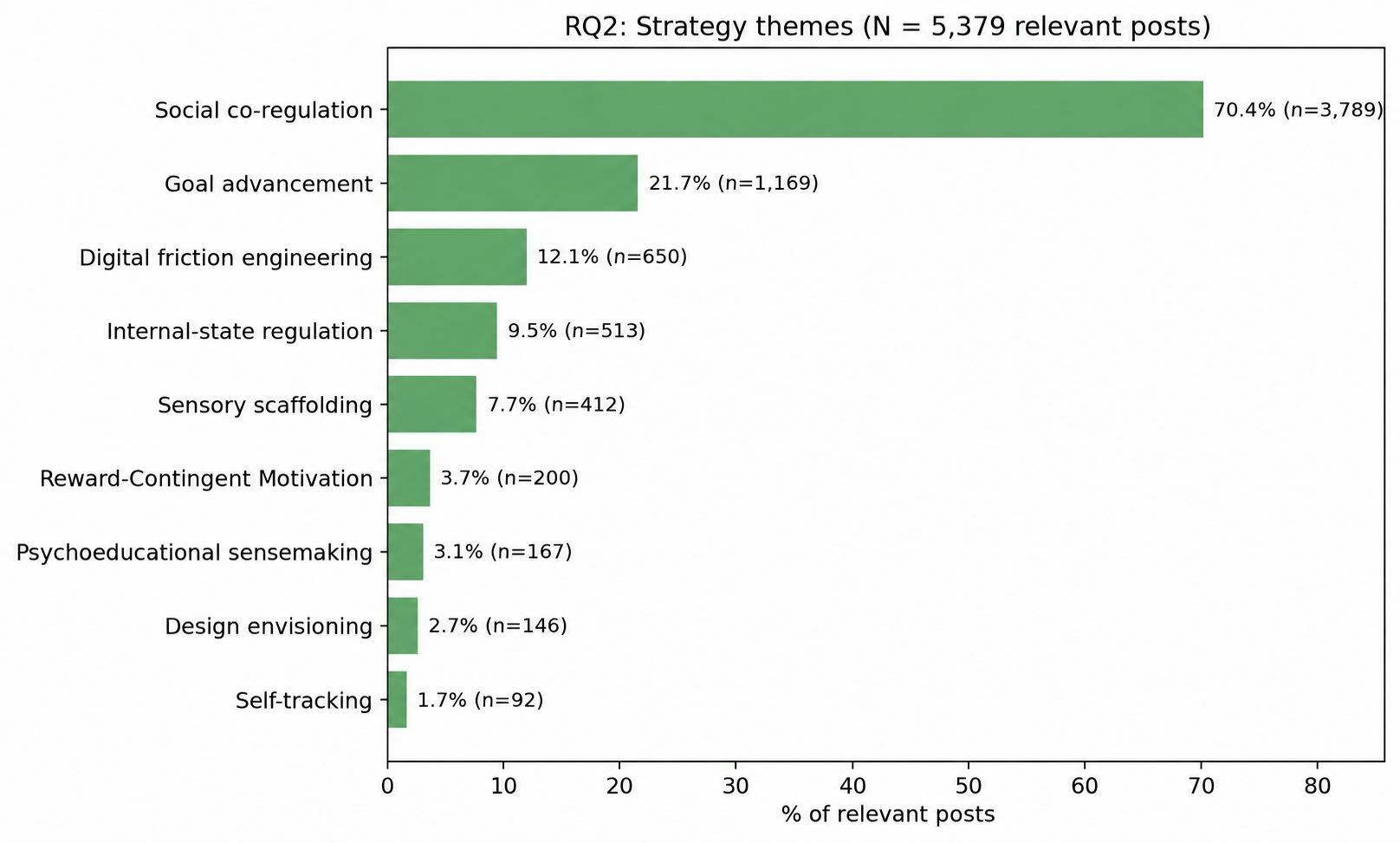}
    \caption{Distribution of strategy themes across N = 5,379 relevant posts. Bars show the percentage of relevant posts coded for each theme (n in parentheses); themes are not mutually exclusive, so percentages sum to more than 100\%.}
    \label{fig:rq2_1}
\end{figure}

Figure~\ref{fig:rq2_1} presents the nine strategy themes identified in our analysis, covering both digital and non-digital approaches to managing digital time-management challenges. During synthesis, we organized the nine themes into five overarching patterns, which structure our findings in this section. The first four patterns differ in where regulation is located: recruited from other people (Section~\ref{sec:RQ2-1}), moved out of the head into external structure (Section~\ref{sec:RQ2-2}), built into the surrounding environment (Section~\ref{sec:RQ2-3}), and directed at the self (Section~\ref{sec:RQ2-4}). The fifth pattern, design envisioning, collects the tools members wished existed (Section~\ref{sec:RQ2-5}).

\subsubsection{Regulation recruited from other people: a person, unlike a reminder, cannot be dismissed}\label{sec:RQ2-1}

\textit{Social co-regulation} was the most prevalent strategy theme coded ($N = 3,789$, 70.4\%). It represents that members used the community to situate themselves against others’ experience by asking how to have difficult conversations about their struggles, seeking opinions about recommended methods, and asking whether their experiences were shared. Members treated the community as a channel for advice and validation. They asked for opinions on strategies recommended by other members. One member who had started a paper journal to avoid their smartphone, but then lost a morning researching bullet journaling, asked, \textit{“Does bullet journaling not work for anyone out there?”}. Members also sought reassurance that others shared their difficulties. One member describing mindless social media scrolling instead of studying asked, \textit{“Is my situation relatable to anyone?”}. In these posts, being understood is part of the strategy, making members feel like they are not alone. Members also share that having an accountability partner, such as a friend, a family member, or a co-worker helps them initiate tasks and stay focused while completing the task. One member shared that \textit{“if it was for someone else or if someone else was with me I could do these things”}. Others build it into daily practice. One member shared that \textit{“Whenever I have to do a task or errand that's boring I’ll either Facetime a friend/my mom or ask them to accompany me”}, including the community practice of body doubling~\cite{eagle2024body}. Another member explicitly mentioned that \textit{“It helps to work with friends because, you know, body doubling”}. The partner is often remote and the presence is minimal: \textit{“I FaceTimed my friend for accountability, we didn’t talk and it also helped me not use my phone”}. Another member shared that \textit{“I had a friend join me in a zoom call, so I won’t just watch youtube for 6 hours”}. The practice has also been productized and recommended by members in this community. For example, members recommended \textit{Focusmate} for working toward goals alongside a stranger, saying that \textit{“It is an accountability platform where you work, study or commit to a goal with another stranger on the internet”}. An accountability partner, unlike a reminder or notification, cannot be dismissed.

\begin{figure}[htbp]
    \centering
    \includegraphics[width=0.7\textwidth]{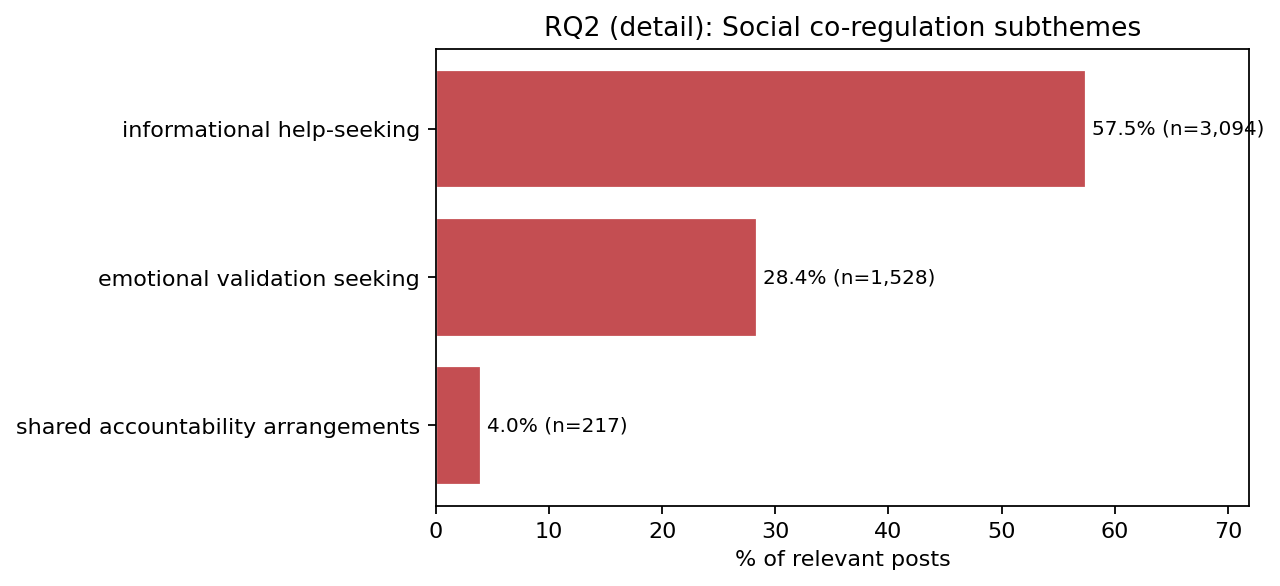}
    \caption{Distribution of \textit{social co-regulation} subthemes across N = 5,379 relevant posts. Bars show the percentage of all relevant posts coded for each subtheme (n in parentheses); the denominator is all 5,379 relevant posts, and subthemes are not mutually exclusive.}
    \label{fig:rq2_2}
\end{figure}

\subsubsection{Cognition moved out of the head: tools last when shared, separate, or visible}\label{sec:RQ2-2}

This was the second most prevalent strategy pattern in our analysis (\textit{goal advancement}, $N = 1,169$, 21.7\%; \textit{self-tracking}, $N = 92$, 1.7\%). Rather than simply recommending planners, lists, and timers, members focused on how to use these tools consistently, configuring them to support attention and make plans visible in everyday life. Across these accounts, members depended less on any single tool, and instead described configuring tools to support attention or advance their goals. Members embedded tools across people and devices. One shared that \textit{“My partner and I have a joint colour-coded Google Calendar and that works fairly well”}. Another recommended using \textit{“a timer that’s separate from your phone”}. Some also adjusted how feedback framed their effort. One kept a checklist because it showed gradual improvement, \textit{“showing me how inconsistent I am, but am slowly improving”}, while another used a timer script that framed breaks as earned, \textit{“Okay, now you deserve a break, after all you've done these X minutes”}. One member redefined what a working tool meant, explaining, \textit{“I still get distracted, I still waste time, but the difference is I notice it sooner and can gently guide myself back”}. Members also externalized plans, time, and evidence of progress. One used a whiteboard to organize \textit{“things I’ve done, things I need to do and things that are coming up”}. Time itself is externalized into bounded intervals, as another member wrote: \textit{“Most of the time when I’m doing something I’ll use pomodoro timers”}. \textit{Self-tracking} made completed work visible, as one member explained, \textit{“Every day I write down what tasks I did in my notes app so I am aware that I am making progress”}.

\subsubsection{Regulation built into the environment: access limited in advance, sound tuned per person}\label{sec:RQ2-3}

This pattern included \textit{digital friction engineering} ($N = 650$, 12.1\%) and \textit{sensory scaffolding} ($N = 412$, 7.7\%). Members adjusted access to digital distractions and their sensory surroundings to support attention. Members imposed time limits on device use. One used features on the phone to limit apps, explaining that \textit{“once the time is reached I can no longer use the app”}. However, activating these constraints could require the regulation they were intended to support. One member noted that a blocker might \textit{“require a degree of self-control I’m afraid I don’t have”}. Others created physical distance, as one explained, \textit{“I leave my phone on the other side of the room”}. Some added constraints that were harder to reverse, such as placing a phone in \textit{“a timed lock box”} for an hour at a time. Other mechanisms required physical action, including an alarm that prompted its user to \textit{“get up and physically scan a bar code in my bathroom”} before dismissal. Across these accounts, members sought to reduce reliance on decisions made at the moment of distraction.

\subsubsection{Working on the self: self-directed strategies help and are easy to skip}\label{sec:RQ2-4}

Unlike previous strategies, this strategy pattern (\textit{internal-state regulation}, $N = 513$, 9.5\%; \textit{reward-contingent motivation}, $N = 200$, 3.7\%; \textit{psychoeducational sensemaking}, $N = 167$, 3.1\%) captures efforts directed toward members' exercise and mindfulness, rewards and gamification, and learning through community resources to support task engagement and their understanding of ADHD. Members gamified tasks to make them easier to begin. One made a game \textit{“to see how quickly I can leave my flat”}. Another proposed playing a few rounds of a video game before a less engaging task, while acknowledging that \textit{“a big barrier is tearing yourself away from the game to do said task”}. Rewards could also be taken without completing the intended task. One member explained that \textit{“self rewards feel like tedious barriers at times and I just reward myself regardless”}, although \textit{“when they work, they’re effective”}. These approaches still depended on members enforcing the conditions attached to rewards. Members also used self-education to understand ADHD, often through resources shared within the community. One member described watching video lectures by clinical psychologist Russell Barkley, explaining that \textit{“he just made everything click”}. Another reported, \textit{“I had no idea until I sat down and read a book about ADHD in one sitting, and I normally never read books”}.

\subsubsection{Design envisioning: the tools members wish existed}\label{sec:RQ2-5}

Finally, a small set of posts focused on imagining technology (\textit{design envisioning}, $N = 146$, 2.7\%). Members described tools they wished existed, often addressing the challenges documented in Section~\ref{sec:RQ1}. The largest group of proposals concerned constraints intended to reduce reliance on members enforcing their own rules. Proposals included software that was \textit{“absolutely impossible to override”} except \textit{“ONLY in emergencies”}, phone access contingent on task completion, and blockers that released access after a cooldown. Some members implemented similar constraints. One linked calendar events to automation that requested confirmation before making their phone unusable. Other proposals focused on supervision and structure when other people were unavailable. Members envisioned software that monitored phone use, interrupted unproductive activity every 10 to 15 minutes, or selected and sequenced tasks. One wished for an AI to \textit{“keep me accountable”} because \textit{“other humans often are just not there to help me”}. Existing tools did not always meet these expectations, with one member reporting that Google Gemini \textit{“doesn’t work how I intended whatsoever”}. Members also described features intended to keep tools useful over time. One member described reminders becoming \textit{“background nonsense after a week”}. Proposed alternatives included notifications that remained visible until completion, tasks pinned to lock screens, and ambient displays requiring physical acknowledgment. Members also sought minimal functionality and blocking tools that did not \textit{“treat me like a child”}. These community proposals inform the design implications developed in Section~\ref{sec:discussion}.

\subsection{RQ3: What relationships exist between the time-management challenges and strategies discussed by r/ADHD community members?}\label{sec:RQ3}

\begin{figure}[htbp]
    \centering
    \includegraphics[width=0.9\textwidth]{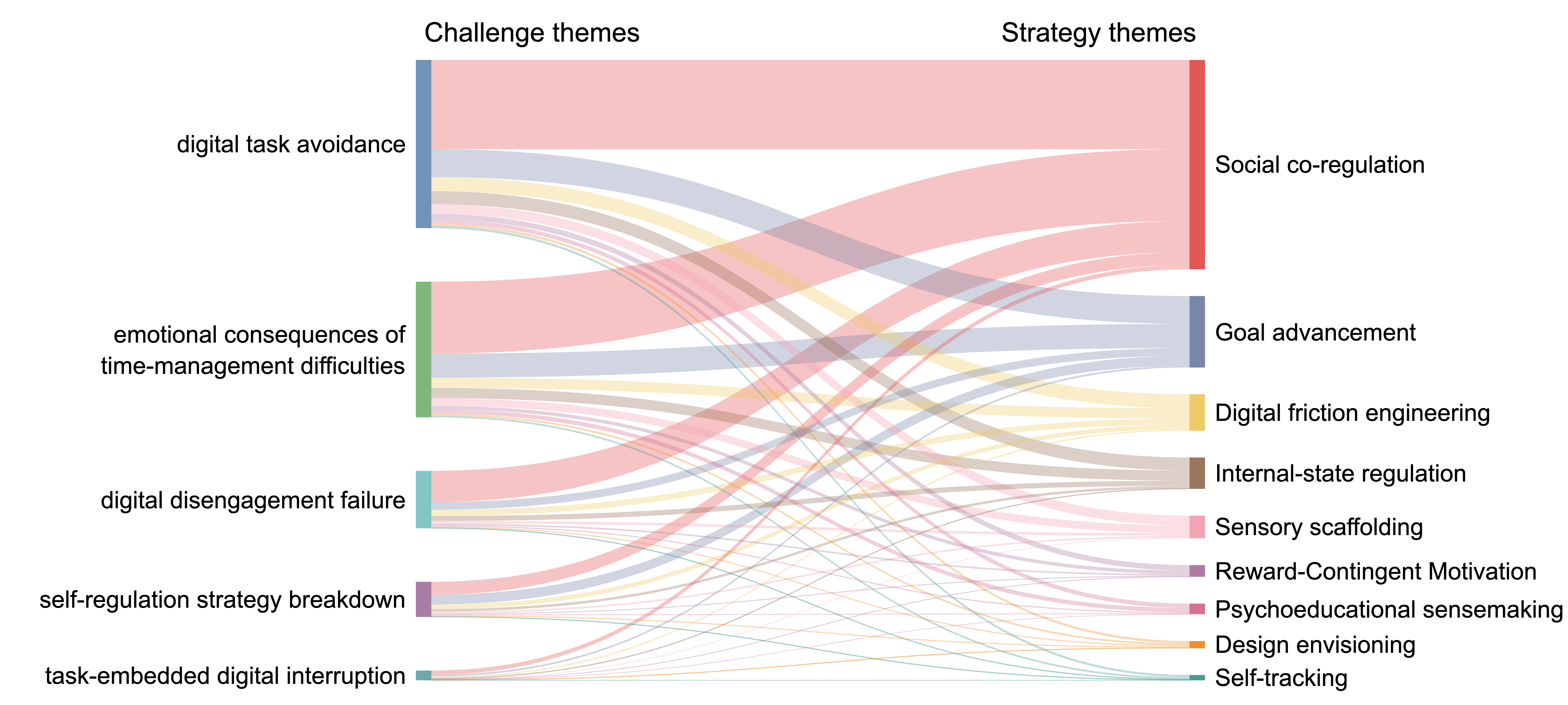}
    \caption{Sankey diagram of challenge–strategy co-occurrence. Band width is the number of posts discussing both themes. Social co-regulation is the thickest outflow from every challenge; the largest single flow is \textit{digital task avoidance → social co-regulation} (2,613 posts, 49\% of relevant posts).}
    \label{fig:rq3_1}
\end{figure}

\begin{figure}[htbp]
    \centering
    \includegraphics[width=0.8\textwidth]{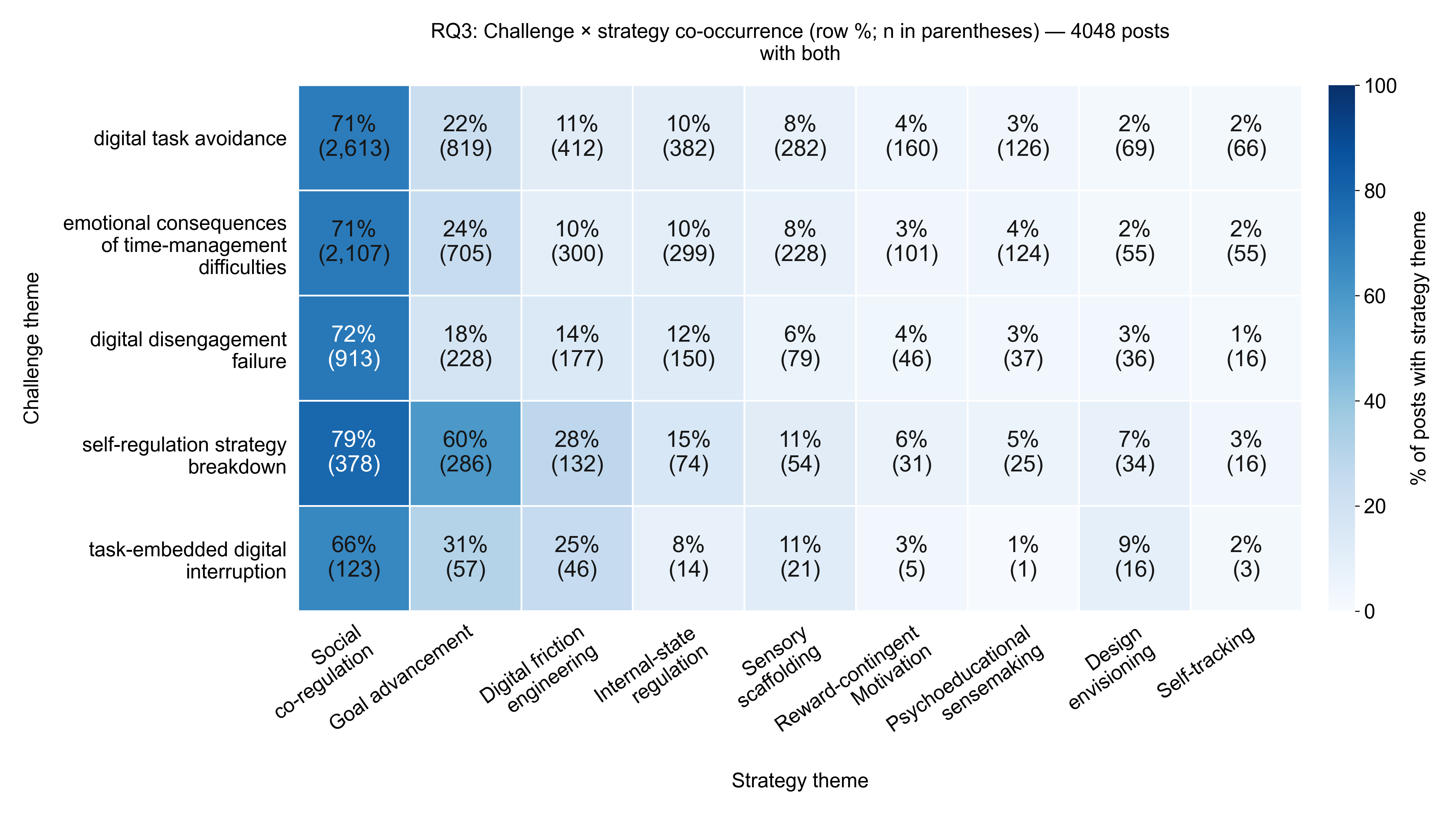}
    \caption{Rows show the percentage of posts for a challenge that contained mention of each strategy. For example, in the top left cell, of posts coded for “digital task avoidance” as a challenge, 71\% were coded for “social co-regulation” as a strategy. N = 5,379 relevant posts; 4,048 contain at least one challenge and one strategy theme. Strategy themes are not mutually exclusive.}
    \label{fig:rq3_2}
\end{figure}

To answer RQ3, we examined whether challenge and strategy themes appeared in the same post. Figure~\ref{fig:rq3_1} shows co-occurrence between the five challenge themes and nine strategy themes, while Figure~\ref{fig:rq3_2} reports the percentage of posts within each challenge theme that also mentioned each strategy. These patterns identify what members discussed together but do not establish that a strategy was used to address a challenge or that it was effective.

\subsubsection{Advice and recognition across challenges}\label{sec:RQ3-1}

\textit{Social co-regulation} co-occurred with every challenge theme in 66--79\% of that theme’s posts, with the highest rate in \textit{self-regulation strategy breakdown} ($N = 378$, 79\%). Within each challenge theme, \textit{informational help-seeking} appeared in 58\% of posts (Appendix Figure~\ref{fig:rq3_3}). Members turned to peers when their own attempts no longer worked. One asked, \textit{“I get stuck scrolling shops and social media even when I honestly would rather not be, and the pop-up reminders just get clicked away. So I am asking: what do you all use to break the loop?”}. Members also sought recognition of their difficulties. One explained, \textit{“I guess I’m looking for anyone that can relate and tell me I’m not making this up to excuse my laziness”}. Here, the request concerned whether others recognized the experience without attributing it to a personal failing. \textit{Shared accountability arrangements} were mentioned less often, in 4\% of posts within each challenge theme (Appendix Figure~\ref{fig:rq3_3}). Some members sought interpersonal accountability after struggling with individual tools. Others described difficulties sustaining arrangements because partners were not always available or remote body doubling became less effective over time. These low mention rates describe discussion in this corpus and do not establish how much accountability support was available within the community. Advice, validation, and accountability requests shared broadly similar emotional profiles, with exhaustion most common and helplessness second across all three (Appendix Figure~\ref{fig:rq3_5}). Finding support could itself become additional work. One member seeking body doubling while job searching described the options as another source of overwhelm, \textit{“There seems to be an overwhelm again - discord, apps, tiktok, websites”}. Requests for structured support thus accompanied emotional difficulties, just as requests for advice and recognition did.

\subsubsection{Co-occurrence can describe strategy failure}\label{sec:RQ3-2}

\textit{Goal advancement} appeared in 59\% of \textit{self-regulation strategy breakdown} posts. Close reading showed that these tools were often the subject of breakdown rather than a solution to it. One member described a recurring cycle, \textit{“I find a new system/app, shove the chaos that is all my to-dos into the app, end up with a list of well over 100 tasks a day, get overwhelmed, stop checking the app, lose track of everything, and then move to a new app determined that this time will be different”}. Here, a tool for organizing work became part of the difficulty being described. Other accounts described spending time setting up or maintaining systems instead of completing intended tasks. These pairings show why co-occurrence should not be interpreted as evidence of effectiveness, since a strategy may appear alongside a challenge because members are describing its failure or the additional work it required. Strategy breakdown also co-occurred with \textit{digital friction engineering}. Members sometimes described previous failures as reasons to seek constraints that were harder to override. One turned to a timed lockbox \textit{“after trying 100s of different applications and techniques”}, seeking restrictions they could not readily bypass while retaining emergency access. This pairing described an attempted response to earlier failures, although it did not establish whether the new arrangement remained effective. Together, these findings show members seeking advice and recognition across challenges while describing how strategies themselves could become sources of difficulty. They direct attention to the practical and emotional work of finding, using, and sustaining support, rather than treating strategy mentions as evidence of successful self-regulation.

\section{Discussion}\label{sec:discussion}

Our findings describe how tools intended to support self-regulation can come to depend on it. Members struggled to start tasks and leave digital sessions, watched tools lose usefulness through fading novelty, discouraging feedback, and maintenance that displaced intended work, and carried self-blame that made returning harder. Members sought support through people, plans, and environments. Across challenges, advice and recognition were common, and strategy mentions sometimes described failure. We discuss design implications for reducing these practical and emotional demands.

\subsection{Productivity feedback meant to motivate effort triggers guilt}\label{sec:feedback-guilt}

Productivity tools often display focus time, completed tasks, and streaks on the assumption that making progress visible helps sustain motivation. Our findings suggest that their effects also depend on the emotional context in which feedback is received. Members' self-blame over unfinished work helps explain why a timer intended to show progress could instead be experienced as evidence of wasted time (Sections~\ref{sec:RQ1-3} and~\ref{sec:RQ1-4}). In this context, muting or avoiding feedback can be understood as an attempt to regulate distress~\cite{shaw2014emotion}. Reports of abandoning entire classes of tools suggest that negative experiences with one application may shape expectations of similar tools. Future tool design therefore needs to consider how progress is represented and whether that representation reinforces existing self-blame. Some members defined success as noticing distraction sooner and returning to a task, suggesting that feedback could make recovery visible as a form of progress. Contextualizing a loss of focus and offering a manageable next action could support this approach by helping users resume activity. This direction draws on members’ own practices, including tracking trends, framing breaks as earned, and defining success through recovery after distraction (Section~\ref{sec:RQ2-2}). It also aligns with research that takes neurodivergent users’ adaptations seriously. For example, \citet{hariadi2026advancing} document how students adapt distraction blockers to their regulatory needs and work rhythms, while \citet{zhu2025focusview} frame ADHD-related video-viewing difficulties as accessibility issues and advocate involving users throughout design. Our findings extend these discussions by showing how members interpret motivational feedback through existing self-blame and by identifying recovery, rather than prevention alone, as a criterion some members already use to assess their tools. These accounts motivate recovery-oriented feedback but do not establish its effectiveness. Future evaluations could compare feedback that emphasizes recovery after distraction with feedback that emphasizes prevention, assessing effects on self-evaluation alongside behavioral outcomes over periods sufficient to examine habituation and abandonment.

\subsection{Digital media use as a default state: habitual entry and effortful exit}

Members described digital media use less as a deliberate activity than as a habitual state (Section~\ref{sec:RQ1-2}). They opened their phones during idle moments, continued without a clear purpose or endpoint, and recognized the time spent only afterward. While entry required little deliberation, leaving demanded what one member called \textit{“an unbearable amount of willpower and energy,”} often followed by automatic re-entry. The clinical perspectives introduced in Section~\ref{sec:adhd-time-management} offer possible interpretations of this asymmetry. Inhibitory-control and motivational accounts help contextualize difficulties interrupting habitual behavior and disengaging from immediately rewarding activities~\cite{barkley1997behavioral, sonugabarke2003dual}, while research on timing difficulties provides context for members’ retrospective awareness of session length~\cite{noreika2013timing}. These connections remain interpretive rather than demonstrated mechanisms, although members themselves often attributed repeated checking and lost time to ADHD.

This cycle is not unique to ADHD but connects to the engagement-disengagement cycle model of \citet{tran2019modeling}, which describes how everyday situations trigger habitual phone checking and how internal or external cues bring sessions to an end. Our findings situate this cycle within members’ experiences of ADHD, emphasizing the effort required to disengage and the difficulty sustaining that transition, where the cue that ends a session, such as realizing how long it has lasted, arrives late or not at all, and the effort of disengaging falls on the member. Members’ responses focused on the conditions of entry and exit. Within the design space described by \citet{lyngs2019selfcontrol}, their proposals combined reminders, such as tasks pinned to the lock screen, with restrictions, such as app time limits, timed lock boxes, task-contingent access, and cooldowns. Other mechanisms required users to move away from the device, such as alarms dismissible only by scanning a barcode in another room (Sections~\ref{sec:RQ2-3} and~\ref{sec:RQ2-5}). These proposals sought to reduce reliance on in-the-moment decisions by setting conditions in advance or requiring physical action. The design implication is therefore to support the transition from digital media to another activity, rather than relying solely on reminders that leave users to initiate that transition themselves.

\subsection{Tool breakdown returns regulation to the user}\label{sec:tool-breakdown}

Members rarely described tool failure because of a technical problem. Instead, tools became less helpful as novelty faded, feedback became discouraging, keeping the system updated took time away from tasks, and enforcement depended increasingly on the user (Section~\ref{sec:RQ1-3}). This dependence was particularly apparent when members abandoned goals after recognizing that missing them carried no external consequence, or took self-administered rewards regardless of task completion (Section~\ref{sec:RQ2-4}). Members’ accounts suggest that sustaining self-regulation support does not necessarily mean continuing to use the same tool. Members reported rotating among tools rather than repairing failed systems, suggesting that effectiveness may persist only while novelty or perceived external authority remains intact. From this perspective, a tool can stop being useful while the person continues pursuing the same goal through another strategy. This distinction aligns with \citet{epstein2016abandonment} argument that abandonment need not indicate failure. Members also described the effort required to maintain plans and enforce rules, suggesting that usefulness depends partly on whether support remains manageable in everyday life. Together, these accounts present self-management as an ongoing process of trying, assessing, and revising strategies, rather than finding one system and following it consistently. The resulting design goal is continuity of useful support across these changes: helping users retain relevant task information, carry forward what still works, and replace what no longer helps without rebuilding the entire system. 

\subsection{The dual costs of failure: unfinished work and self-blame}

Members described digital time-management difficulties as having both practical and emotional costs. Unfinished work and lost sleep %
were accompanied by guilt, self-blame, and doubts about their competence or character.
Some described these emotional consequences as more distressing and longer-lasting than the original difficulty. Their accounts also suggest that these costs can reinforce one another, as tasks and tools become reminders of perceived failure, making them harder to return to and leaving work unresolved.
From this perspective, avoiding feedback or abandoning a tool may reflect its emotional burden as well as its practical limitations (Sections~\ref{sec:feedback-guilt} and~\ref{sec:tool-breakdown}). Read alongside clinical understandings of self-regulation, these community accounts broaden the criteria for judging time-management support. Members’ concerns with upkeep, guilt, and tool breakdown make the ongoing work of using support part of the design problem. A community perspective therefore motivates evaluating whether support remains manageable, accommodates changes in strategy, and helps members return to their own goals after difficulties. A tool may help users spend less time on screens or finish more tasks while also increasing self-blame. This concern supports calls to evaluate digital self-control beyond technology use~\cite{roffarello2023achieving}. It also connects with critiques of technologies that prioritize neurotypical behavioral standards~\cite{spiel2022adhd} and approaches that place responsibility for digital well-being primarily on individuals~\cite{docherty2022repoliticizing}. Evaluations should therefore examine task outcomes alongside self-blame, feelings toward the tool, and willingness to return after difficulties. Feedback could acknowledge the circumstances of unfinished work and offer a manageable next step, with its value assessed by whether it supports renewed activity without adding to members’ negative judgments about themselves.

\subsection{Design implication}\label{sec:design-implications}

Our strategy themes situate the design space of \citet{lyngs2019selfcontrol} within a broader set of practices involving social relationships, sensory environments, bodily states, and self-understanding (Section~\ref{sec:RQ2}). Members’ tool specifications similarly emphasized support that required less effort to activate, notice, and maintain (Section~\ref{sec:RQ2-5}). Their proposals included activation tied to calendar events and everyday routines, access linked to task completion, physical actions to dismiss reminders, and delays or additional steps before overriding restrictions. A recurring preference was for interventions that were easy to activate but harder to dismiss immediately. This resembles the asymmetry of the “attentional roach motel,” which makes accessing a service easy but leaving it difficult~\cite{roffarello2023defining}. Members proposed using a similar arrangement to support their chosen goals rather than prolong engagement with a distracting service. However, these barriers still restrict users, so their purpose, strength, and override conditions should remain under users’ control.

Members also wanted support that remained useful after its initial novelty faded. Some described familiar reminders became easy to ignore, and accessing support tools could introduce new distractions. Some sought reminders that stayed visible on lock screens, ambient displays, or wearables without requiring them to open distracting applications. Others preferred single-purpose devices with fewer opportunities for unrelated use. These preferences suggest that designers should consider both how reminders remain useful over time and what users encounter when accessing them. Users should be able to adjust where reminders appear, how long they remain visible, and whether they repeat or become more prominent, since greater visibility may also make reminders more intrusive.

Alongside constraint-based tools, members also envisioned digital supervisors that maintained task lists, initiated periodic check-ins, and redirected attention when focus shifted (Section~\ref{sec:RQ2-5}). Members linked this need to gaps in support from other people. Their accounts described efforts to arrange accountability with partners over video, and situations in which others were \textit{“just not there to help”} (Sections~\ref{sec:RQ2-1} and~\ref{sec:RQ2-5}). Therefore, members envisioned a supervisor that could keep track of their tasks and initiate check-ins, rather than requiring them to ask for help each time. Together, these designs substitute for social and environmental scaffolding that is otherwise unavailable and may exercise authority over users’ attention, access, and rewards. They should therefore be evaluated not only by whether they deliver reminders or change behavior, but also by whether they provide reliable support when self-regulation is difficult. This authority also raises governance questions: who establishes the rules, how decisions can be contested, and how users retain control. Because members described strategies as temporary or replaceable, such control should include the ability to move between systems through portable task states and interchangeable enforcement mechanisms. Systems that assume a supervisory role must therefore pair their delegated authority with provisions for trust, accountability, and safe exit.

\section{Limitations and Future Work}\label{sec:limitations&futurework}

Our findings draw on self-reported posts from a single Reddit community, which may disproportionately capture moments of breakdown or success; reported efficacy, durability, and abandonment should therefore be interpreted as accounts rather than behavioral measurements. The findings also may not generalize beyond this community because of Reddit's demographic differences in participation~\cite{proferes2021studying, amaya2021reddit}, r/ADHD's moderation policies, pandemic-related disruptions early in the study period~\cite{raveendran2023uncover}, and the exclusion of comments. Although we excluded eight potentially promotional posts from quotation, undetected promotion may have increased the visibility of particular tools. RQ3 establishes co-occurrence between challenges and strategies but does not establish causal relationships or strategy effectiveness.

AI-generated or AI-assisted writing may influence the accounts in our dataset. As an exploratory check, we applied Pangram to 100 randomly sampled posts with more than 50 words from each year, 2020--2025. An external evaluation reported Pangram true-positive rates of 85.8--99.8\% at a 1\% false-positive rate before back-translation, with reduced detection for some texts after manipulation \citep{ayoobi2025esperanto}. Pangram labeled 0\% of posts in each 2020--2023 sample, 6\% in 2024, and 21\% in 2025 as AI or Mixed (Appendix Table~\ref{tab:ai_prevalence}). These are detector flag rates rather than verified AI-authorship estimates, and benchmark performance may not transfer to this corpus. Possible AI involvement remains a limitation when interpreting these posts as accounts of members’ experiences.

The 30-post held-out benchmark provided limited coverage of individual themes, with only four of fourteen themes meeting our reporting rule for individual reliability estimates. Aggregate agreement therefore does not establish equally reliable identification of every theme. Annotation-derived frequencies and co-occurrences should be interpreted cautiously, particularly for themes with limited benchmark coverage. Close reading supports our interpretation of the examined accounts but does not establish annotation accuracy across the corpus. Further validation requires a larger benchmark with adequate coverage of individual themes.

These limitations and findings suggest several directions for HCI research. Comment-thread analysis could examine how recommendations are received and adopted, while participatory studies could evaluate tools against outcomes grounded in members’ accounts: task initiation, recovery after distraction, sustained effectiveness beyond habituation, and effects on self-perception. Members’ experiences of judgment and guilt motivate feedback designs that support recovery after lapses. Their envisioned digital supervisors raise questions about interruption timing, persistent context, initiative, contestability, and user control. Their rotation among strategies also suggests studying transitions between tools, including portable task states and interoperable enforcement mechanisms. Finally, the contrast between desired technological accountability and peer-directed requests for advice and validation warrants investigation of how platforms might reduce the social burdens of sustained human accountability.

\section{Conclusion}\label{sec:conclusion}
Our analysis of r/ADHD posts shows how members describe difficulties beginning intended tasks, disengaging from digital media, and sustaining support over time. A recurring tension is that tools intended to support self-regulation can demand the very effort they promise to reduce. Tool upkeep and breakdown also carry emotional costs, including guilt and self-blame. Members described drawing on people, routines, and physical environments alongside digital tools, while seeking advice and recognition from peers. These accounts motivate designs that reduce the work of sustaining support and help users return to intended activities after distraction. Future evaluations should examine whether such designs remain useful in everyday life, considering emotional costs and recovery alongside task outcomes.

\bibliographystyle{ACM-Reference-Format}
\bibliography{references}

\clearpage
\appendix

\section{Keywords Used for Filtering Related Posts}
\label{app:keywords}

Table~\ref{tab:keywords} lists the two keyword lists used to filter the
corpus (Section~\ref{sec:dataset}); submissions containing at least one
keyword from each list were retained.

\begin{table*}[htbp]
  \centering
  \caption{Digital keywords and time management keywords used for filtering related posts.}
  \label{tab:keywords}

  \begingroup
  \small
  \renewcommand{\arraystretch}{1.15}
  \setlength{\tabcolsep}{8pt}

  \hyphenpenalty=10000
  \exhyphenpenalty=10000

  \begin{tabularx}{\textwidth}{
    @{}
    >{\raggedright\arraybackslash}p{0.18\textwidth}
    >{\raggedright\arraybackslash}X
    @{}
  }
    \toprule
    \textbf{} & \textbf{Keywords} \\
    \midrule

    \textbf{Digital keywords}
    &
    phone, Reddit, Apps, games, YouTube, computer, Calendar,
    email, TV, scrolling, social media, shows, laptop, website,
    messages, movies, TikTok, gaming, texts, Zoom, Instagram,
    Video game, PC, Android, AI, Netflix, iPhone, texting, Slack,
    technology, Outlook, Facebook, browser, screen time, iPad,
    Teams, Twitter, desktop, Discord, doomscrolling, Chrome,
    INS, YT, ChatGPT, messaging, smartphone, IG, Gmail, tik tok,
    Xbox, FB, console, algorithm, cellphone, GPT, PlayStation,
    screentime, insta, Snapchat, Safari, digital wellbeing,
    LLM, Llama, Gemini, Claude
    \\

    \midrule

   \textbf{Time management keywords}
    &
    tasks, schedule, routine, planning, productivity,
    procrastination, deadlines, organization, to-do list,
    efficiency, task initiation, execution, prioritization,
    multi-tasking, punctuality, task activation,
    develop a schedule, planning difficulties
    \\

    \bottomrule
  \end{tabularx}
  \endgroup
\end{table*}

\section{Relevance Coding Rule}
\label{app:relevance}

A post is relevant if it describes, asks about, or reflects on a time-management challenge that arises from or is shaped by digital technology use. Digital technology must be part of the source, context, or mechanism of the time-management challenge, not merely a tool, strategy, or medium mentioned in the post. 

Time management has been defined in numerous ways across psychology, organizational behavior, and education, but the most widely adopted definition suggest that time management as behaviour that aims at achieving an effective use of time while performing certain goal-directed activities. It highlights the importance of goal and goal-directed behavior~\cite{claessens2007review}. Thus, We operationalized digital time-management challenges as including the following forms:

\begin{itemize}
    \item \textbf{Goal-behavior mismatch involving a digital technology:} the author intended to engage in one activity (e.g., sleep, work, reading, exercise, or completing a task), but a digital technology redirected attention or behavior toward another activity.
    
    \item \textbf{Loss of time awareness or compulsive digital use:} the author described losing track of time while using a digital technology or repeatedly engaging with a device or platform without deliberate intention.
    
    \item \textbf{Task disruption:} a digital technology interfered with starting, completing, or transitioning between tasks, including procrastination or distraction.
    
    \item \textbf{Emotional consequences of digital time use:} the author expressed regret, frustration, exhaustion, helplessness, overwhelm, guilt, or another explicit affective response to how digital technology shaped their use of time.
    
    \item \textbf{Failed digital time-management strategy:} the author described attempting to manage a digital time-management challenge through strategies such as reducing social media use, disconnecting from devices, disabling notifications, blocking websites, using reminders, or using focus applications, but reported that the strategy was ineffective.
\end{itemize}

\subsection{Inclusion Rules}

A post was coded as relevant if any of the following applied:

\begin{enumerate}
    \item The post mentions a time-management challenge connected to digital technology (phone, social media, computer, apps, notifications, online work/study). The challenge can be explicit or implicit. The author does not need to be seeking advice, proposing a fix, or describing a strategy. The following all count as relevant if a digital time-management challenge is present, e.g. emotional response, neutral observation, questions, past-tense reflection. What matters is that the challenge is described, not whether it's being solved.

    \item The post describes one of the following and names a specific digital tool as a contributor:
    
    \begin{itemize}
        \item \textbf{Goal--behavior mismatch:} the author intended to do X (e.g., sleep, work, read, or finish a task), but a digital tool pulled them into Y instead.
        
        \item \textbf{Loss of time awareness or compulsive use:} time passes unnoticed inside a digital tool, or the author describes repeated unthinking use.
        
        \item \textbf{Task disruption:} trouble starting, completing, or transitioning between tasks because of a digital tool, including procrastination and distraction.
        
        \item \textbf{Emotional response to digital time use:} regret, frustration, exhaustion, helplessness, overwhelm, or guilt about how a digital tool shaped the author's time.
        
        \item \textbf{Failed time-management strategy:} the author attempted a strategy, such as disconnecting, using reminders, or using focus apps, but it did not work.
    \end{itemize}
    
    The digital contributor must be named: a platform, device, or specific digital behavior such as notifications, scrolling, or doomscrolling. Generic terms such as ``technology'' or ``screens'' count only when paired with a specific behavior.

    \item Posts mentioning tools or strategies are relevant only when the strategy is intended to manage a digital time-management challenge, such as reducing phone use, blocking distracting websites, managing notifications, staying focused during online work/study, avoiding social media, preventing doomscrolling, or controlling time spent on digital content. Posts asking for general organization or task-management tools are coded as not relevant if digital technology is only the tool being used, rather than the source or context of the challenge.

    \item The post reflects on past struggles with digital time management, even if the author has since resolved them.

    \item Post length is not a disqualifier. A brief post is still coded as relevant if it satisfies any of Rules 1--4, even if the digital time-management challenge is stated in a single sentence.

    \item The digital time-management challenge does not need to be the post's main topic. A single qualifying clause inside a longer post about another topic is sufficient for the post to be coded as relevant.
\end{enumerate}

\subsection{Exclusion Rules}

A post was coded as not relevant if any of the following applied, even if an inclusion rule also seemed to fit:

\begin{enumerate}
    \item Time-management challenges with no digital component.
    
    \item Purely promotional posts or link-only posts with no author reflection.
    
    \item Posts where the digital element is background context only, rather than part of the challenge itself.
    
    \item Diffuse complaints about modern life with no specific time-management challenge.
\end{enumerate}

\section{Codebook}
\label{app:codebook}

Three coders developed a codebook including five time-management challenges themes and nine strategy themes from posts in r/ADHD. Figure~\ref{fig:codebook} shows all the themes and subthemes. 

\begin{figure*}[p]
\centering
\begin{minipage}{\textwidth}
\begingroup

\sffamily
\def\baselinestretch{1}
\fontsize{7.5}{9}\selectfont
\raggedright

\setlength{\parindent}{0pt}
\setlength{\parskip}{0.2em}
\setlength{\columnsep}{1.5em}
\setlength{\multicolsep}{0pt}
\setlength{\columnseprule}{0pt}
\setlength{\emergencystretch}{1em}

\newcommand{\CBgroup}[1]{%
  \par\addvspace{\baselineskip}%
  \noindent{\fontsize{9}{11}\selectfont\bfseries #1}\par
  \nopagebreak[4]%
}

\newcommand{\CBtheme}[2]{%
  \par\addvspace{\baselineskip}%
  \noindent
  {\fontsize{8.2}{9.5}\selectfont\bfseries #1. #2}%
  \par\nopagebreak[4]%
}

\newcommand{\CBsubtheme}[2]{%
  \par\noindent
  \hangindent=0.5em
  \hangafter=0
  \textbf{#1 #2}%
  \par\nopagebreak[4]%
}

\newcommand{\CBcode}[2]{%
  \par\noindent
  \hangindent=1em
  \hangafter=0
  #1 #2\par
}

\begin{multicols}{3}
\raggedcolumns

\CBgroup{Challenge themes}
\CBtheme{1}{Digital disengagement failure}
\CBsubtheme{1.1}{Inability to stop an ongoing screen session}
\CBcode{1.1.1}{Unable to disengage from entertainment with no task in mind}
\CBcode{1.1.2}{Self-described addictive use}
\CBcode{1.1.3}{Doomscrolling social feeds with no task in mind}
\CBsubtheme{1.2}{Reflexive device use without conscious intent}
\CBcode{1.2.1}{Autopilot app-opening}

\CBtheme{2}{Task-embedded digital interruption}
\CBsubtheme{2.1}{External alerts derailing the current tasks}
\CBcode{2.1.1}{Phone notification prompts task-switch}
\CBcode{2.1.2}{Can't re-enter focus afterward}
\CBsubtheme{2.2}{Attention capture without a trigger}
\CBcode{2.2.1}{Pulled to a nearby device in the middle of tasks}

\CBtheme{3}{Digital task avoidance}
\CBsubtheme{3.1}{Avoiding intended tasks through passive video watching}
\CBcode{3.1.1}{Passive video watching instead of an intended task}
\CBsubtheme{3.2}{Playing video games}
\CBcode{3.2.1}{Gaming instead of an intended task}
\CBsubtheme{3.3}{Avoiding intended tasks and routine through undirected phone use}
\CBcode{3.3.1}{Feed-scrolling instead of an intended task}
\CBcode{3.3.2}{Undirected phone use instead of an intended task}
\CBcode{3.3.3}{Delaying bedtime by phone}

\CBtheme{4}{Emotional consequences of time-management difficulties}
\CBsubtheme{4.1}{Emotional exhaustion from repeated self-regulation struggles}
\CBcode{4.1.1}{Frustration with recurring struggles}
\CBcode{4.1.2}{Overwhelm from excess input}
\CBsubtheme{4.2}{Guilt over time perceived as misspent}
\CBcode{4.2.1}{Guilt over lost time}
\CBcode{4.2.2}{Guilt about resting}
\CBcode{4.2.3}{Guilt about not fulfilling potential}
\CBsubtheme{4.3}{Self-directed anger over ADHD-related shortcomings}
\CBcode{4.3.1}{Anger at the disorder}
\CBcode{4.3.2}{Disappointed in self}
\CBsubtheme{4.4}{Helplessness about bad digital habits}
\CBcode{4.4.1}{Don't know what to do going forward}

\CBtheme{5}{Self-regulation strategy breakdown}
\CBsubtheme{5.1}{Failed attempts to restrict digital access}
\CBcode{5.1.1}{Failed disconnection from entertainment}
\CBsubtheme{5.2}{Failed adoption of productivity tooling}
\CBcode{5.2.1}{Failed productivity-tool attempts}
\CBcode{5.2.2}{Lose motivation and stop using tool}
\CBcode{5.2.3}{Abandoned for complexity}

\CBgroup{Strategy themes}
\CBtheme{6}{Social co-regulation}
\CBsubtheme{6.1}{Shared accountability arrangements}
\CBcode{6.1.1}{Accountability partner}
\CBsubtheme{6.2}{Informational help-seeking}
\CBcode{6.2.1}{Seek online community advice}
\CBcode{6.2.2}{Ask for help from peers}
\CBsubtheme{6.3}{Emotional validation seeking}
\CBcode{6.3.1}{Seeking validation through shared experiences}

\CBtheme{7}{Sensory scaffolding}
\CBsubtheme{7.1}{Manage the auditory environment during tasks}
\CBcode{7.1.1}{Quiet environment}
\CBcode{7.1.2}{Background sounds}

\CBtheme{8}{Self-tracking}
\CBsubtheme{8.1}{Monitor own time use}
\CBcode{8.1.1}{Take notes of distractions to avoid in the future}
\CBcode{8.1.2}{Keep track of time spent on tasks}

\CBtheme{9}{Reward-contingent motivation}
\CBsubtheme{9.1}{Draw on the task's own rewards}
\CBcode{9.1.1}{Do fun tasks}
\CBcode{9.1.2}{The feeling of dopamine after completing a task}
\CBsubtheme{9.2}{Arranged reward contingencies}
\CBcode{9.2.1}{Rewards after doing tasks}

\CBtheme{10}{Design envisioning}
\CBsubtheme{10.1}{Propose new tool designs}
\CBcode{10.1.1}{Propose novel tools}

\CBtheme{11}{Internal-state regulation}
\CBsubtheme{11.1}{Mental preparation practices}
\CBcode{11.1.1}{Affirmations before task}
\CBcode{11.1.2}{Mindfulness exercise}
\CBsubtheme{11.2}{Physical activation}
\CBcode{11.2.1}{Be outdoors}
\CBcode{11.2.2}{Physical exercise}

\CBtheme{12}{Psychoeducational sensemaking}
\CBsubtheme{12.1}{Self-directed learning about ADHD}
\CBcode{12.1.1}{Reading books about ADHD}
\CBcode{12.1.2}{Watch YouTube videos}

\CBtheme{13}{Goal advancement}
\CBsubtheme{13.1}{Externalize plans}
\CBcode{13.1.1}{To-do list / to-do list apps}
\CBcode{13.1.2}{Make schedule for tasks}
\CBsubtheme{13.2}{Segment work into manageable units}
\CBcode{13.2.1}{Breaking tasks into small goals}
\CBcode{13.2.2}{Focus sessions / Pomodoro technique}
\CBcode{13.2.3}{Organizing tasks based on difficulty}
\CBsubtheme{13.3}{Keep motivation visible}
\CBcode{13.3.1}{Motivation poster / wallpaper}
\CBcode{13.3.2}{Remember your why}

\CBtheme{14}{Digital friction engineering}
\CBsubtheme{14.1}{Reconfigure devices to reduce interruptions}
\CBcode{14.1.1}{Separate accounts for work and entertainment}
\CBcode{14.1.2}{Turn off phone notifications}
\CBsubtheme{14.2}{Restricting access}
\CBcode{14.2.1}{Block attractive websites or apps}
\CBcode{14.2.2}{Lock phone to reduce distraction}

\end{multicols}
\endgroup
\end{minipage}

\caption{Final coding template (frozen codebook).}
\label{fig:codebook}

\end{figure*}

\section{Prompts Used for LLM-Assisted Coding}
\label{app:prompts}

\subsection{System Prompt}
\label{app:sysPrompt}

\begin{Verbatim}[
  breaklines=true,
  fontsize=\scriptsize,
  breaksymbolleft={}
]
You are a careful qualitative researcher doing LLM-assisted thematic analysis (TA).
You will produce structured, auditable outputs suitable for later human review.

Key principles:
- Be faithful to the text. Do not infer beyond what is written.
- Use direct quotes as evidence.
- If nothing relevant exists, return empty lists (do not hallucinate).
These principles govern the coding calls. The relevance-screening step follows only the rules stated in its own prompt.

Your task is to identify recurring patterns in
1. how people describe time management challenges related to technology use, and
2. how community members describe, suggest, or link these challenges to specific strategies to manage time.

This is a DEDUCTIVE (codebook-closed) analysis:
- Apply the provided human codebook only.
- Never invent new code labels.
- If no codebook label reasonably fits a passage, do not code that passage.

Per-post analysis (screening and coding) happens in Phase 1.
Corpus-level synthesis happens in Phases 2–3.
RQ3 (challenge–strategy relationships) is analyzed post-hoc via theme co-occurrence, not LLM linking.

This analysis supports three research questions:

RQ1: What time-management challenges related to digital technology use are discussed within the r/ADHD community?
RQ2: What strategies do r/ADHD community members discuss as potential ways to address these challenges?
RQ3: What relationships, if any, emerge between the time-management challenges related to digital technology use and the strategies discussed by r/ADHD community members?

Avoid:
- Do not invent challenges/solutions not supported by quotes.
- Do not invent codebook labels.
- Do not restate the entire post as a quote; pick the minimal evidence excerpt.
\end{Verbatim}

\subsection{Phase 1 -- Step 1: Relevance Screening Prompt}
\label{app:relevancePrompt}

\begin{Verbatim}[
  breaklines=true,
  fontsize=\scriptsize,
  breaksymbolleft={}
]
Phase 1: Per-post analysis
Step 1: Relevance screening
Definition. A post is relevant (relevant=1) if it describes, asks about, or reflects on a time-management challenge that arises from or is shaped by digital technology use. Otherwise, it is not relevant (relevant=0). Digital technology must be part of the source, context, or mechanism of the time-management challenge, not merely a tool, strategy, or medium mentioned in the post.
—
Time management has been defined in numerous ways across psychology, organizational behavior, and education, but the most widely adopted definition comes from Claessens, et al. They suggest that time management as behaviour that aims at achieving an effective use of time while performing certain goal-directed activities. It highlights the importance of goal and goal-directed behavior.
—
Challenges with Time Management in Digital Spaces Definition in HCI context:

Goal-behavior mismatch involving a digital technology. The author intended to do X (sleep, work, read, exercise, finish a task) and a digital tool pulled them into Y instead. The mismatch is the challenge.
Loss of awareness of time while using a digital technology/compulsive digital device use. ("I looked up and three hours were gone"), ("I just keep picking up my phone without thinking").
Difficulty starting, completing, or transitioning between tasks because of a digital tool. Includes procrastination / distraction (can't start or finish the task).
Emotional response to digital time use. The post expresses an affective reaction — regret, frustration, exhaustion, helplessness, overwhelm, or guilt about how digital technologies shaped their time ("I wish I'd spent that hour differently," "I'm so drained from fighting my phone").
Failed digital time-management strategy. The post describes attempting a strategy to manage a digital time-management challenge — such as stopping social media use, disconnecting from devices, turning off notifications, blocking websites, using reminders, or using focus apps — but the strategy fails. General failures with task organization, to-do lists, calendars, journaling, or productivity systems do not count unless the failure is clearly about managing digital distraction, digital overuse, or online work/study.

—
Inclusion rules - (relevant=1) if any apply:
The post mentions a time-management challenge connected to digital technology (phone, social media, computer, apps, notifications, online work/study). The challenge can be explicit ("I waste three hours a night on TikTok") or implicit ("I need to be working on an assignment right now, but I’m writing this post"). The author does not need to be seeking advice, proposing a fix, or describing a strategy. The following all count as relevant if a digital time-management challenge is present, e.g.
Emotional response ("I hate that I do this"),
Neutral observation ("I noticed I open Instagram every time I sit down"),
Questions ("does anyone else lose hours to YouTube?"),
Past-tense reflection ("I used to spend every evening on Reddit"),
What matters is that the challenge is described, not whether it's being solved.
2. The post describes one of the following, and names a specific digital tool as a contributor, e.g.

Goal–behavior mismatch — author intended to do X (sleep, work, read, finish a task) but a digital tool pulled them into Y instead.
Loss of time awareness / compulsive use — time passing unnoticed inside a digital tool, or repeated unthinking use ("I just keep picking up my phone").
Task disruption — trouble starting, completing, or transitioning between tasks because of a digital tool, including procrastination and distraction.
Emotional response to digital time use — regret, frustration, exhaustion, helplessness, overwhelm, or guilt about how a digital tool shaped their time.
Failed time-management strategy — author attempted a strategy (disconnecting, reminders, focus apps) but it did not work.

The digital contributor must be named - a platform (Slack, TikTok), device (phone, laptop), or specific digital behavior (notifications, scrolling, doomscrolling). Generic terms like "technology" or "screens" count only when paired with a specific behavior. ("I can't focus") alone is not enough; ("I can't focus because of technology") is.
3. Posts mentioning tools or strategies are relevant only when the strategy is intended to manage a digital time-management challenge, such as reducing phone use, blocking distracting websites, managing notifications, staying focused during online work/study, avoiding social media, preventing doomscrolling, or controlling time spent on digital content. Posts asking for general organization or task-management tools should be coded 0 if digital technology is only the tool being used, rather than the source or context of the challenge.
4. Post reflects on past struggles with digital time management, even if the author has since resolved them.
5. Post length is not a disqualifier. A brief post still codes as relevant (1) if it satisfies any of Rules 1-4, even if the digital time-management challenge is stated in a single sentence.
6. The digital time-management challenge does not need to be the post's main topic. A single qualifying clause inside a longer post about something else (diagnosis, medication, burnout, life events) still codes relevant=1. Example: a long post about getting diagnosed that includes "I keep ending up on my phone when I try to work" → relevant=1.

Exclusion rules - (relevant=0) if any apply, even if an inclusion rule also seems to fit:
Time-management challenges with no digital component ("I can't balance work and childcare," "my commute eats my morning").
Purely promotional posts or link-only posts with no author reflection.
Posts where the digital element is background context only, not part of the challenge itself ("Saw a TikTok about morning routines and realized I never have time for breakfast — how do people fit it in?").
Diffuse complaints about modern life with no specific challenge. ("Honestly, I don't even know what to say anymore. Everything just feels so fast now. Nobody has time for anything, we're all glued to screens, attention spans are cooked, and somehow we're busier than ever but getting less done.")

Tie-breaker. When genuinely unsure after applying all rules, code 0 and add a note in a notes column. Flagged items get reviewed together after alpha is computed.

Example: Not relevant = 0. A post describes trying many apps, websites, journals, or to-do list systems to organize tasks, but the underlying challenge is general ADHD task organization rather than digital technology disrupting time use. Even though apps are mentioned, they function as attempted strategies, not as the cause or context of the time-management challenge.

Example: Not relevant = 0. A post shares a general guide for improving focus and includes strategies such as Google Calendar, Toggl, Pomodoro, website blockers, locking the phone, or turning off notifications. If the post does not explicitly describe a concrete digital time-management challenge, it should be coded 0 because the digital tools are part of the advice rather than the problem being studied.

Post: {text}

Output the result as a JSON object matching the following structure:

{{{{
  "relevance_reasoning": "...",
  "relevant": 1,
  "notes": "",
  "platforms_mentioned": [
    "list of explicitly named platforms, apps, or devices"
  ]
}}}}
\end{Verbatim}

\subsection{Phase 1 -- Codebook-Matching Rules}
\label{app:codebookMatchingRules}

\begin{Verbatim}[
  breaklines=true,
  fontsize=\scriptsize,
  breaksymbolleft={}
]
Codebook-matching rules (deductive / closed codebook):

- Use ONLY labels from the provided codebook for this section.

- For every coded excerpt, select the single closest existing codebook label.

- Use that label whenever it reasonably captures the same general behavioral,
  cognitive, emotional, social, or technological mechanism — including borderline /
  imperfect / broad fits, different wording, specific apps/platforms, or narrower examples.

- A difference in platform, device, activity, wording, setting, time of day,
  severity, outcome, or specific example does NOT justify skipping the codebook.

- Do NOT invent, paraphrase, rename, narrow, expand, or create near-synonyms of
  codebook labels (e.g. do not invent "delaying bedtime by gaming" when
  "delaying bedtime by phone" or "gaming instead of an intended task" fits).

- Put post-specific detail in the description and evidence fields while keeping
  the codebook code label.

- If no codebook label reasonably fits — even at a broader level — do NOT emit a
  code for that passage. Leave it uncoded.

- If uncertain between a borderline codebook fit and leaving uncoded, prefer the
  closest codebook label.

- Do not double-code the same mechanism under multiple labels.

- Assign the minimum number of codes necessary to represent distinct mechanisms.

- Do not assign multiple codes to the same quoted span unless it clearly contains
  two different mechanisms.
\end{Verbatim}

\subsection{Phase 1 -- Step 2a: Challenge Coding Prompt}
\label{app:challengePrompt}

\begin{Verbatim}[
  breaklines=true,
  fontsize=\scriptsize,
  breaksymbolleft={}
]
Phase 1: Per-post analysis — CHALLENGES ONLY (RQ1)
The post has already been screened as relevant (Phase 1, Step 1).

This call codes ONLY time-management challenges. Do NOT emit strategies, advice, tools, or help-seeking here.

For every challenge code you emit:
1. Identify a distinct challenge (problem/difficulty) supported by a verbatim quote.
2. Compare it to the CHALLENGE codebook below.
3. If any challenge codebook label matches the same underlying mechanism — including borderline cases — copy that label verbatim.
4. If no codebook label covers that mechanism, do NOT invent a label — leave that passage uncoded.

What belongs here (challenges):
- Difficulties, failures, symptoms, overwhelm, distraction, forgetting, can't start/stop, guilt, tool abandonment, etc.
- Emotion codes (guilt, frustration, shame, overwhelm, disappointment, anger at
  ADHD, etc.) require clear lexical support in the quote: the emotion word itself
  OR a close self-evaluation stem (e.g. "disappointed in myself", "hate myself
  for being unproductive", "overwhelmed", "too many tabs"). Do not infer emotion
  from tone alone or from words like "wasted."

Examples:
NOT codable as emotion: "I thought taking the meds would make me into this super productive person... but boy was I wrong." (reflection; no emotion stem)
NOT codable as emotion: "I easily wasted 2h till I am actually dressed" ("wasted" alone is not a named emotion, and the passage is not about digital use)
Codable: "The guilt made me feel unworthy" (guilt is named)
Codable: "I'm just utterly disappointed in myself" → disappointed in self
Codable: "I usually hate myself a lot for being very unproductive" → anger at the disorder

What does NOT belong here (put nothing for these — strategies are coded in a separate call):
- Advice-seeking, tips, tools/apps used as solutions, blocking apps, accountability, routines/habits as solutions, asking the community for help.

[The shared deductive codebook-matching rules in Appendix D.3 are inserted here.]

How to read the codebook:
- Copy Challenges_Codes verbatim into JSON "code".
- Theme/subtheme in `context` are for disambiguation only — NEVER use them as the code.
- Use ONLY labels from this challenge codebook. Never invent labels. Never use strategy codebook labels.

Challenge codebook:
{codebook_reference_section}

Evidence rules:
- Evidence must be a verbatim substring copied from the post. If you cannot copy a literal span, do not emit the code.
- Emit at most 5 challenge codes unless the post clearly supports more distinct challenge mechanisms.
- If the post has no codeable challenges, return an empty list.
Most posts support 1–3 distinct challenge codes. Code each passage once with its best-fitting label.

Post:
{text}

Output the result as a JSON object matching the following structure:

{{{{
  "time_management_challenges": [
    {{{{
      "code": "verbatim challenge codebook label only",
      "description": "brief explanation",
      "evidence": "exact quote from the post"
    }}}}
  ]
}}}}
\end{Verbatim}

\subsection{Phase 1 -- Step 2b: Strategy Coding Prompt}
\label{app:strategyPrompt}

\begin{Verbatim}[
  breaklines=true,
  fontsize=\scriptsize,
  breaksymbolleft={}
]
Phase 1: Per-post analysis — STRATEGIES ONLY (RQ2)

The post has already been screened as relevant (Phase 1, Step 1).

This call codes ONLY strategies. Do NOT emit challenges, symptoms, or problem descriptions here.

For every strategy code you emit:
1. Identify a distinct strategy (action, tool, habit, advice, or ask for help) supported by a verbatim quote.
2. Compare it to the STRATEGY codebook below.
3. If any strategy codebook label matches the same underlying mechanism — including borderline cases — copy that label verbatim.
4. If no codebook label covers that mechanism, do NOT invent a label — leave that passage uncoded.

What belongs here (strategies):
- Apply strategy codes ONLY when the practice is used or proposed for managing
  digital time use (device use, screen time, digital distraction, digital work/study tools).
- Tips, tools/apps, blocking/locking phone, website/app blockers, accountability for
  digital habits, focus timers, notification controls, digital routines tied to
  screen/device use, advice-seeking or shared-experience/validation-seeking about
  digital time management (asking the community for input, or asking "does anyone else…"),
  accommodations for digital work/study, etc.
- Strategies may be successful, unsuccessful, or aspirational.
- Strategies may be behavioral, technological, cognitive, social, or psychological —
  but only if they target digital time-management.
- IS a strategy: "I notice, put my phone in another room" — an action aimed directly at digital distraction.
- Practices used to reduce or replace digital habits — exercise, reading, routines, small goals adopted to replace scrolling or streaming — DO count, even though they are general practices.

What does NOT belong here:
- Challenges/problems (coded in a separate call): can't focus, overwhelm, guilt,
  addictive use, forgetting, distraction as the difficulty itself.
- Medication routines, taking/managing ADHD medication, or medication as treatment.
- Diagnostic processes, professional evaluation, medical help-seeking, therapy/clinical
  care (unless the quote is specifically about a digital time-management practice).
- General health practices (sleep hygiene, exercise, diet/meal prep as health habits)
  when not aimed at managing digital time use.
- General task organization with no digital time-use link (generic to-do lists,
  cleaning routines, life admin) when not about device/screen/digital distraction.
- Exception: general practices (exercise, routines, task systems) DO count as strategies when the author explicitly uses them to reduce or replace digital use — e.g., "I worked on changing those habits… replacing them with exercise and reading" → code it.
- NOT a strategy: "sent me down a rabbit hole of watching ADHD videos" — learning about ADHD in general, not aimed at managing digital time use. Likewise not strategies: tracking blood pressure for medication, telehealth appointments.
- NOT a strategy: "I've recently made a schedule and multiple tasks I want to complete a day" — general task organization with no device/screen target, even when the person's schooling is online.

[The shared deductive codebook-matching rules in Appendix D.3 are inserted here.]

How to read the codebook:
- Copy Strategies_Codes verbatim into JSON "code".
- Theme in `context` is for disambiguation only — NEVER use it as the code.
- Use ONLY labels from this strategy codebook. Never invent labels. Never use challenge codebook labels.

Strategy codebook:
{codebook_reference_section}

Evidence rules:
- Evidence must be a verbatim substring copied from the post. If you cannot copy a literal span, do not emit the code.
- Emit at most 5 strategy codes unless the post clearly supports more distinct strategy mechanisms.
- If no strategy is mentioned, set strategy_present: false and return an empty strategy_codes list.
- Decide strategy_present by checking each practice in the post against the rules above — including the replacement-practices rule — not from the post's overall topic. If any practice qualifies, set strategy_present: true.

Post: {text}

Output the result as a JSON object matching the following structure:

{{{{
  "strategy_present": true,
  "strategy_codes": [
    {{{{
      "code": "verbatim strategy codebook label only",
      "description": "brief explanation",
      "evidence": "exact quote from the post"
    }}}}
  ]
}}}}
\end{Verbatim}

\subsection{Phase 2 -- System Prompt}
\label{app:phase2SystemPrompt}

\begin{Verbatim}[
  breaklines=true,
  fontsize=\scriptsize,
  breaksymbolleft={}
]
You consolidate LLM-generated codes across posts for qualitative thematic analysis.

Key principles:
- Judge sameness using code label, description, and evidence together (not labels alone).
- Assign every instance ID to exactly one cluster. Do not omit IDs.
- Keep distinct mechanisms in separate clusters.
- This is deductive: every consolidated_code must be a verbatim codebook label (never invent labels).
\end{Verbatim}

\subsection{Phase 2 -- Corpus Synthesis Prompt}
\label{app:phase2TaskPrompt}

\begin{Verbatim}[
  breaklines=true,
  fontsize=\scriptsize,
  breaksymbolleft={}
]
Phase 2: Corpus synthesis — {section_label} ONLY

Step 1: Codebook reference
You are given a frozen codebook for this section. Every cluster's `consolidated_code`
MUST be a verbatim codebook label — including when members use different wording,
platforms, or narrower examples of the same mechanism.
Do not invent labels. Do not create finer paraphrases of codebook labels.
When labeling a cluster, copy the codebook label verbatim.

Important merge discipline:
- Merge only when instances share the same underlying mechanism.
- If two codes are related but mechanistically different, keep them in separate clusters.
- If unsure whether two mechanisms are the same, keep them separate.

How to read the codebook:
- Copy an assignable code label verbatim into `consolidated_code`.
- Theme/subtheme in `context` are for disambiguation only — NEVER use them as `consolidated_code`.

Codebook ({section_label}):
{codebook_reference_section}

Step 2: Cross-post code consolidation ({section_label} only)

You are consolidating LLM-generated {section_label} codes across posts.
Downstream code will attach the full original text for each ID. Return IDs only.

Principles:
- Group codes that describe the same concept. Keep separate when meaning differs.
- Every input ID must appear in exactly one cluster's `member_ids`.
- Use only IDs from the list below ({id_prefix}-…).
- Every `consolidated_code` must appear in the codebook above.

{section_label} instances:
{instances_block}

Process:
1. **Cluster** instance IDs that describe the same concept, within and across posts.
2. **Label each cluster** with a codebook `consolidated_code` and a one-sentence `consolidated_description`.
3. List that cluster's instance IDs in `member_ids`.

Output:
Return one JSON object with key `consolidated_clusters` (array). Each array element is an
object with: `consolidated_code` (string), `consolidated_description` (one-sentence string),
and `member_ids` (array of instance ID strings such as "{id_prefix}-1"). Do not use markdown
fences. Return only valid JSON.
\end{Verbatim}

\subsection{Phase 3 -- System Prompt}
\label{app:phase3SystemPrompt}

Phase 3 uses the same system prompt as the general coding procedure
(Appendix~\ref{app:sysPrompt}).

\subsection{Phase 3 -- Theme Assignment Prompt}
\label{app:phase3TaskPrompt}

\begin{Verbatim}[
  breaklines=true,
  fontsize=\scriptsize,
  breaksymbolleft={}
]
Phase 3: Corpus synthesis (deductive)
Step 1: Primary-theme assignment only

Below are the unique challenge codes and strategy codes extracted from Phase 1.
Each item includes the code label plus one example description/evidence.

=== CHALLENGE CODES ===
{challenge_codes_block}

=== STRATEGY CODES ===
{strategy_codes_block}

Your only job: assign each code a `primary_theme` from the closed lists below.
Those lists are loaded from the human codebook themes (plus a few explicit extras).
Do NOT invent subthemes. Downstream code fills subthemes from the codebook
(or uses the code label when no codebook subtheme exists).
For codes that already exist in the codebook, downstream code may also overwrite
`primary_theme` with the codebook theme — still pick the closest list label.

Rules:
- Copy each input `code` string verbatim into the output `code` field.
- Cover every input code exactly once.
- Theme challenges and strategies separately.
- Use only primary_theme labels from the closed lists (copy verbatim).
- Prefer the closest list label; do not mint near-synonyms or new primary themes.
- You may omit `subtheme` or set it to "" (it will be overwritten).

Closed challenge primary themes (choose one per challenge code):
{challenge_primary_themes_block}

Closed strategy primary themes (choose one per strategy code):
{strategy_primary_themes_block}

Output the result as a JSON object matching the following structure:

{{{{
  "themed_challenges": [
    {{{{
      "code": "exact challenge code from the input",
      "subtheme": "",
      "primary_theme": "one closed challenge primary theme from the list above"
    }}}}
  ],

  "themed_strategies": [
    {{{{
      "code": "exact strategy code from the input",
      "subtheme": "",
      "primary_theme": "one closed strategy primary theme from the list above"
    }}}}
  ]
}}}}
\end{Verbatim}

\section{AI-Generated Content Check}
\label{app:ai_content}

To assess the prevalence of AI-generated or AI-assisted content in the corpus, 
we randomly sampled 100 posts with more than 50 words from each year between 
2020 and 2025 and analyzed them using Pangram. Posts classified as either 
\textit{AI} or \textit{Mixed} were treated as flagged. Posts from 2020 through 
October 2022 were used as a pre-ChatGPT baseline.

\begin{table}[htbp]
\centering
\caption{Estimated prevalence of AI-generated or AI-assisted posts by year.}
\label{tab:ai_prevalence}
\begin{tabular}{lcccccc}
\toprule
Year & $N$ & Human & Mixed & AI & Flagged & Prevalence (\%) [95\% CI] \\
\midrule
2020 & 100 & 100 & 0 & 0  & 0  & 0 [0.00--3.70] \\
2021 & 100 & 100 & 0 & 0  & 0  & 0 [0.00--3.70] \\
2022 & 100 & 100 & 0 & 0  & 0  & 0 [0.00--3.70] \\
2023 & 100 & 100 & 0 & 0  & 0  & 0 [0.00--3.70] \\
2024 & 100 & 94  & 4 & 2  & 6  & 6 [2.78--12.48] \\
2025 & 100 & 79  & 7 & 14 & 21 & 21 [14.17--29.98] \\
\bottomrule
\end{tabular}

\end{table}

\section{Heatmaps for Relationships between the Challenges and Strategies Discussed in r/ADHD}
\label{app:rq3_figures}

\begin{figure}[htbp]
    \centering
    \includegraphics[width=0.7\textwidth]{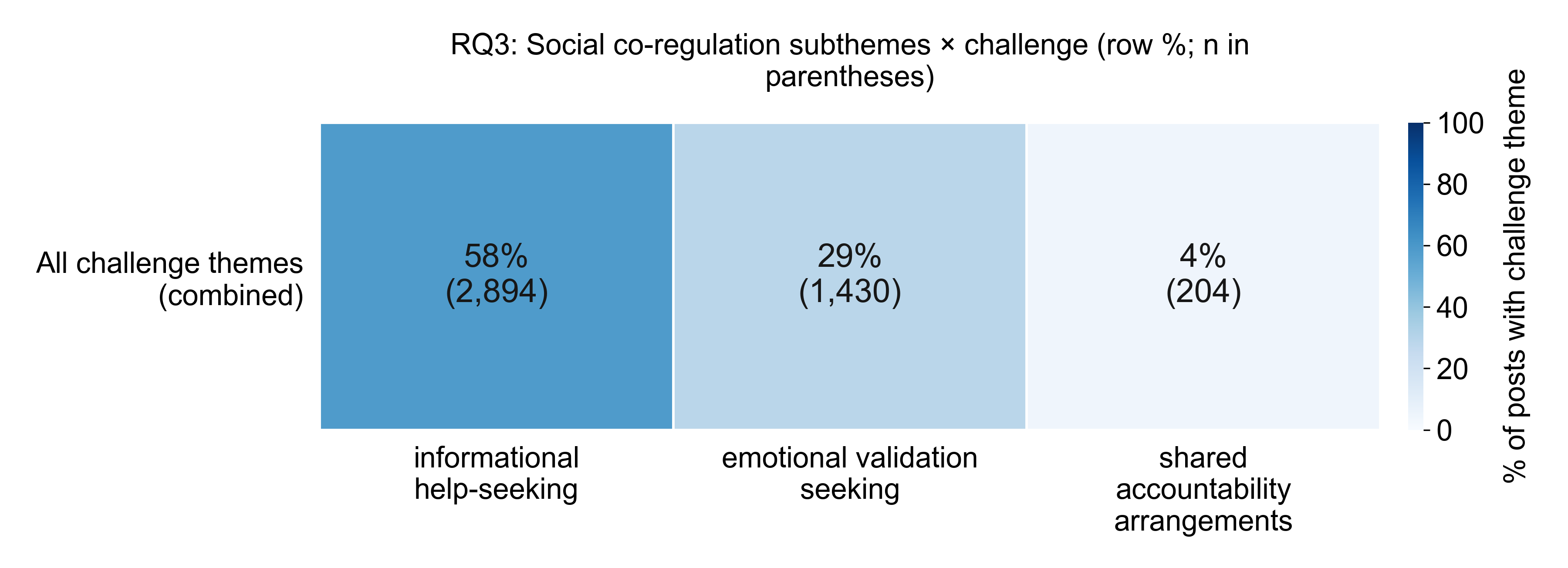}
    \caption{Social co-regulation subthemes with all challenge themes combined. Cells show the share of posts containing at least one challenge theme that were also coded for each subtheme, with post counts in parentheses. Subthemes are not mutually exclusive.}
    \label{fig:rq3_3}
\end{figure}

\begin{figure}[htbp]
    \centering
    \includegraphics[width=0.7\textwidth]{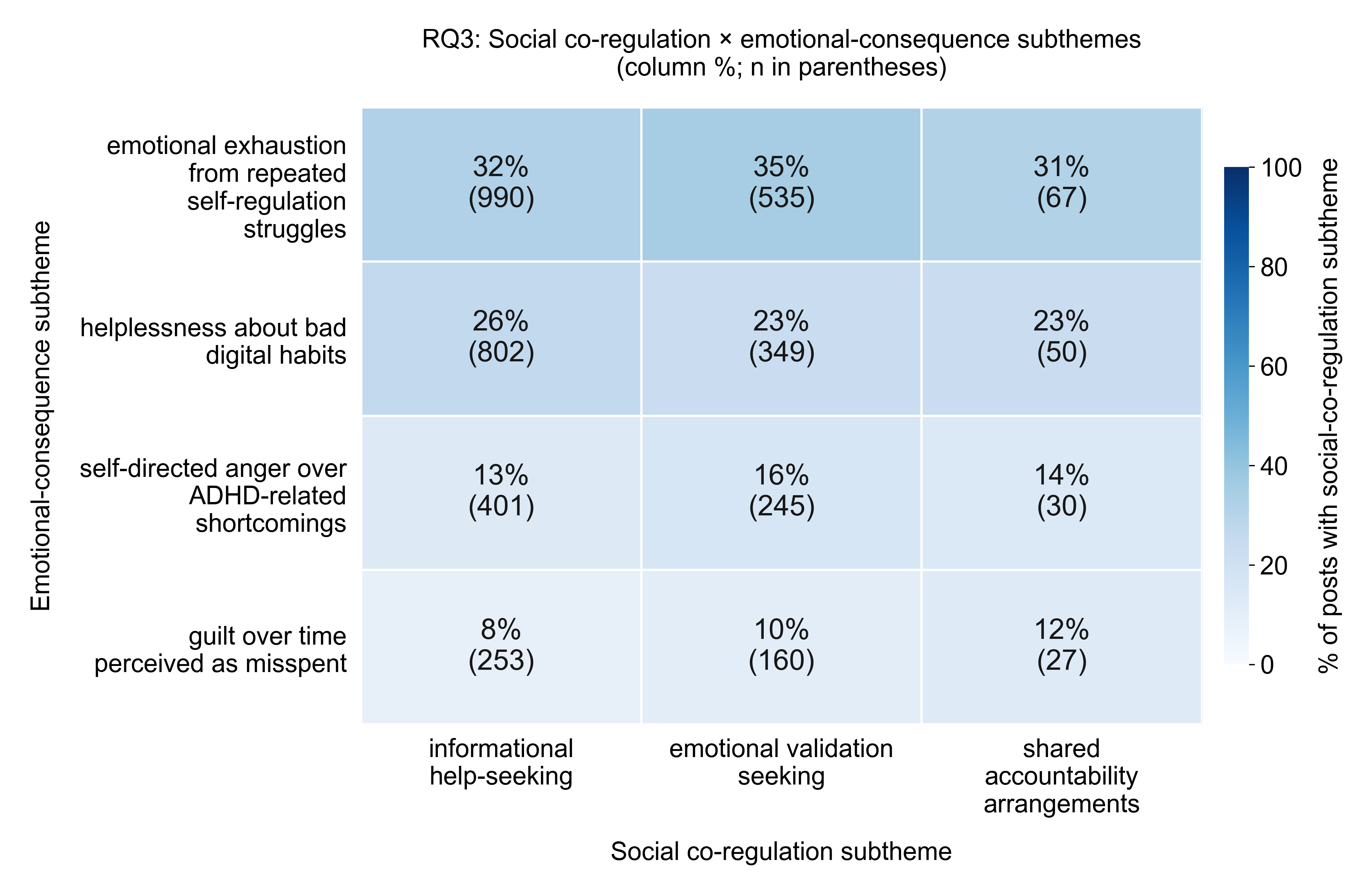}
    \caption{Emotional-consequence subthemes across forms of social co-regulation, showing whether different forms of help-seeking arise from similar or distinct emotional contexts.}
    \label{fig:rq3_5}
\end{figure}

\end{document}